\documentclass[fleqn,usenatbib]{mnras}

\usepackage[utf8]{inputenc}

\usepackage[T1]{fontenc}
\usepackage{ae,aecompl}

\usepackage{graphicx}	
\usepackage{amsmath}	
\usepackage{amssymb}	
\usepackage{wasysym}
\usepackage[varg]{txfonts}
\usepackage{subcaption}
\usepackage{booktabs}
\usepackage{natbib,twoopt}
\usepackage{tabularx}
\usepackage{color}
\usepackage{adjustbox}
\usepackage[flushleft]{threeparttable}

\newcommand{\ts}{\textsubscript}
\newcolumntype{M}[1]{>{\centering\arraybackslash}m{#1}}

\title[Individual optical variability of AGNs from MEXSAS2]{Individual optical variability of Active Galactic Nuclei\\ from the MEXSAS2 sample\thanks{Full Table \ref{tab:cds} is only available at the CDS via anonymous ftp to \url{http://cdsarc.u-strasbg.fr/} (130.79.128.5).} }

\author[M. Laurenti et al.]{
M. Laurenti,$^{1,2}$\thanks{E-mail: \href{mailto:marco.laurenti@roma2.infn.it}{marco.laurenti@roma2.infn.it}}
F. Vagnetti,$^{1,2}$
R. Middei$^{2,3}$
and M. Paolillo$^{4,5,6}$
\\
$^{1}$Dipartimento di Fisica, Università di Roma "Tor Vergata", via della Ricerca Scientifica 1, 00133 Roma, Italy\\
$^{2}$INAF - Osservatorio Astronomico di Roma, via Frascati 33, 00040 Monte Porzio Catone, Italy\\
$^{3}$Space Science Data Center, SSDC, ASI, via del Politecnico snc, 00133 Roma, Italy\\
$^{4}$Dipartimento di Fisica, Università di Napoli "Federico II", via Cinthia 9, 80126 Napoli, Italy\\
$^{5}$INFN - Sezione di Napoli, via Cinthia 9, 80126 Napoli, Italy\\ 
$^{6}$INAF - Osservatorio Astronomico di Capodimonte, via Moiariello 16, 80131 Napoli, Italy
}

\date{Accepted XXX. Received YYY; in original form ZZZ}

\pubyear{2020}

\begin{document}
\label{firstpage}
\pagerange{\pageref{firstpage}--\pageref{lastpage}}
\maketitle

\begin{abstract}
{At present, most of the variability studies of active galactic nuclei (AGNs) are based on ensemble analyses. Nevertheless, it is interesting to provide estimates of the individual variability properties of each AGN, in order to relate them with intrinsic physical quantities. A useful dataset is provided by the Catalina Surveys Data Release 2 (CSDR2), which encompasses almost a decade of photometric measurements of $\sim500$ million objects repeatedly observed hundreds of times.}
{We aim to investigate the individual optical variability properties of 795 AGNs originally included in the Multi-Epoch XMM Serendipitous AGN Sample 2 (MEXSAS2). Our goals consist in: (i) searching for correlations between variability and AGN physical quantities; (ii) extending our knowledge of the variability features of MEXSAS2 from the X-ray to the optical.}
{We use the structure function (SF) to analyse AGN flux variations. We model the SF as a power-law, $\text{SF}(\tau)=A\,(\tau/\tau_0)^\gamma$, and we compute its variability parameters. We introduce the V-correction as a simple tool to correctly quantify the amount of variability in the rest frame of each source.}
{We find a significant decrease of variability amplitude with increasing bolometric, optical and X-ray luminosity. We obtain the indication of an intrinsically weak positive correlation between variability amplitude and redshift, $z$.
Variability amplitude also appears to be positively correlated with $\alpha\ts{ox}$.
The slope of the power-law SF, $\gamma$, is weakly correlated with the bolometric luminosity $L\ts{bol}$ and/or with the black hole mass $M\ts{BH}$.
When comparing optical to X-ray variability properties, we find that X-ray variability amplitude is approximately the same for those AGNs with larger or smaller variability amplitude in the optical.
On the contrary, AGNs with steeper SF in the optical do present steeper SF in the X-ray, and vice versa.}{}
\end{abstract}

\begin{keywords}
surveys -- galaxies: active -- quasars: general 
\end{keywords}



\section{Introduction}
\label{sec:intro}
A key feature of active galactic nuclei (AGNs) consists in their stochastic, aperiodic continuum variability.
Flux variations have been ubiquitously witnessed across the entire range of wavelengths contributing to the spectral energy distribution (SED) of AGNs, although with different amplitudes and timescales.  
In the optical/UV band, these variations are of the order of few tenths of magnitude on the timescale of years \citep[e.g.][]{devries2003, vandenberk2004}.   
In the X-ray band, variability may occur on timescales as short as hours but it is also observed to increase up to at least a few years \cite[e.g.][]{markowitz2004, shemmer2014}.

It is a common thought that different components of AGNs contribute to flux variations in different electromagnetic bands.
While optical/UV variability is probably driven by phenomena which take place in the accretion disk, X-ray variability is usually addressed to physical processes occurring in a hot corona close to the central black hole \citep[BH; e.g.][]{haar93}.
The mechanisms which drive AGN variability are still not completely understood. In the recent years several scenarios have been proposed, such as a single coherent oscillator \citep[e.g.][]{almaini2000}, a superposition of flares/spots due to accretion disk instabilities \citep[e.g.][]{trevese2002, pereyra2006}, starbursts in the host galaxy \citep[e.g.][]{aretxaga1997}, variable absorption and/or reflection \citep[e.g.][]{chevallier2006} and also gravitational microlensing by compact foreground objects \citep[e.g.][]{zackrisson2003}. 

\noindent The importance of variability studies of AGNs is crucial, since they can give insights into the physical properties of the emitting regions.
Different tools are used to analyse the variability of these sources and one of the most popular is the structure function (SF). 
The SF has been widely adopted for ensemble variability studies in the optical/UV band \citep[e.g.][]{bauer2009, kozlowski2016, caplar2017} and also in the X-ray \cite[e.g.][]{vagnetti2011, vagnetti2016, middei2017}. On the contrary, it has been rarely applied to individual variability analyses.

This is due to the fact that such individual studies would require a large number of observations of the same sources at many different epochs.
At present, however, only few surveys can provide a statistically consistent amount of photometric data related to the repeated monitoring of several objects, which are necessary to perform a meaningful individual variability analysis. 
This justifies our choice of using data from the Catalina Surveys \citep[e.g.][]{drake2014}, in order to investigate the individual variability properties of a given AGN sample.

In addition, we are interested in searching for possible correlations between these properties and different characteristic physical quantities of AGNs.
In the following we will always refer to variability amplitude with the term variability, though a number of authors have also investigated characteristic variability timescales \citep[e.g.][]{collier2001, kelly2009, macleod2012}.
According to several studies, variability appears to be negatively correlated with luminosity. This evidence is confirmed in the optical as well as in the X-ray band \citep[e.g.][]{lawrence1993, netzer1996, manners2002, zuo2012, paolillo2017}.
Another parameter whose correlation with variability is often studied, is the black hole mass. For instance, \citet{wold2007} found a positive correlation with this quantity. \citet{wilhite2008} and \citet{macleod2010} did also recover similar results, proposing that the above dependence could be explained in terms of an anti-correlation with the Eddington ratio.
In these terms, \citet{simm2016} could only report an anti-correlation of variability amplitude with the Eddington ratio, while they did not observe any correlation with the black hole mass.
The dependence on redshift is still source of controversial results. Some authors \citep[e.g.][]{li2018} recover a positive correlation, while others find a negative correlation \citep[e.g.][]{cristiani1990, cristiani1996}.
In some other works, variability is not correlated with $z$ at all \citep[e.g.][]{macleod2010}. 
We know that, in every flux-limited sample, luminosity is strongly correlated with redshift.
Hence, it is difficult to completely disentangle this effect while analysing the relation between variability and $z$.
Additionally, in the optical band, a positive correlation between the above quantities could be driven by a frequency effect.
Indeed for a given photometric filter, higher redshift AGNs are observed at bluer rest frame frequencies, where – in the optical – they are usually found to be more variable \citep[e.g.][]{trevese2002, ruan2014, yang2018}.

The paper is organised as follows. Section \ref{sec:data} describes the AGN sample used in our analysis. 
Section \ref{sec:sf} illustrates the computation of the SF along with its results.
In Section \ref{sec:corr} we search for correlations between variability parameters and physical quantities of AGNs.
Section \ref{sec:compX} aims to compare optical to X-ray variability properties of our AGN sample.
In Section \ref{sec:summ} we summarise and discuss the main results of the analysis.

The standard $\Lambda$CDM cosmology ($H_0=70$ km s$^{-1}$ Mpc$^{-1}$, $\Omega_\text{m}=0.3$, $\Omega_\Lambda=0.7$) is adopted throughout the paper.

\begin{figure*}
\vspace{-1cm}
\centering
\includegraphics[width=\columnwidth]{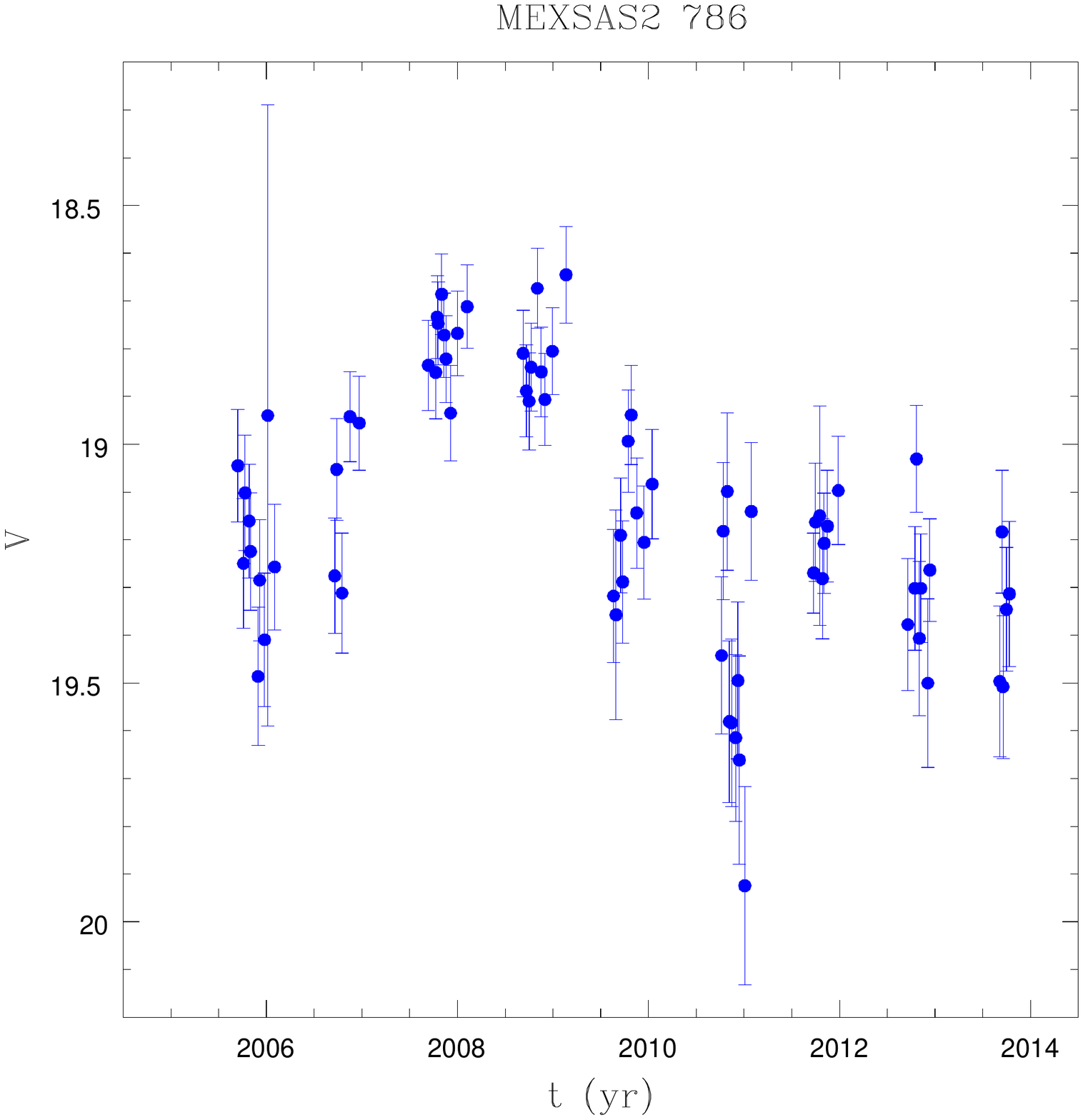}
\includegraphics[width=\columnwidth]{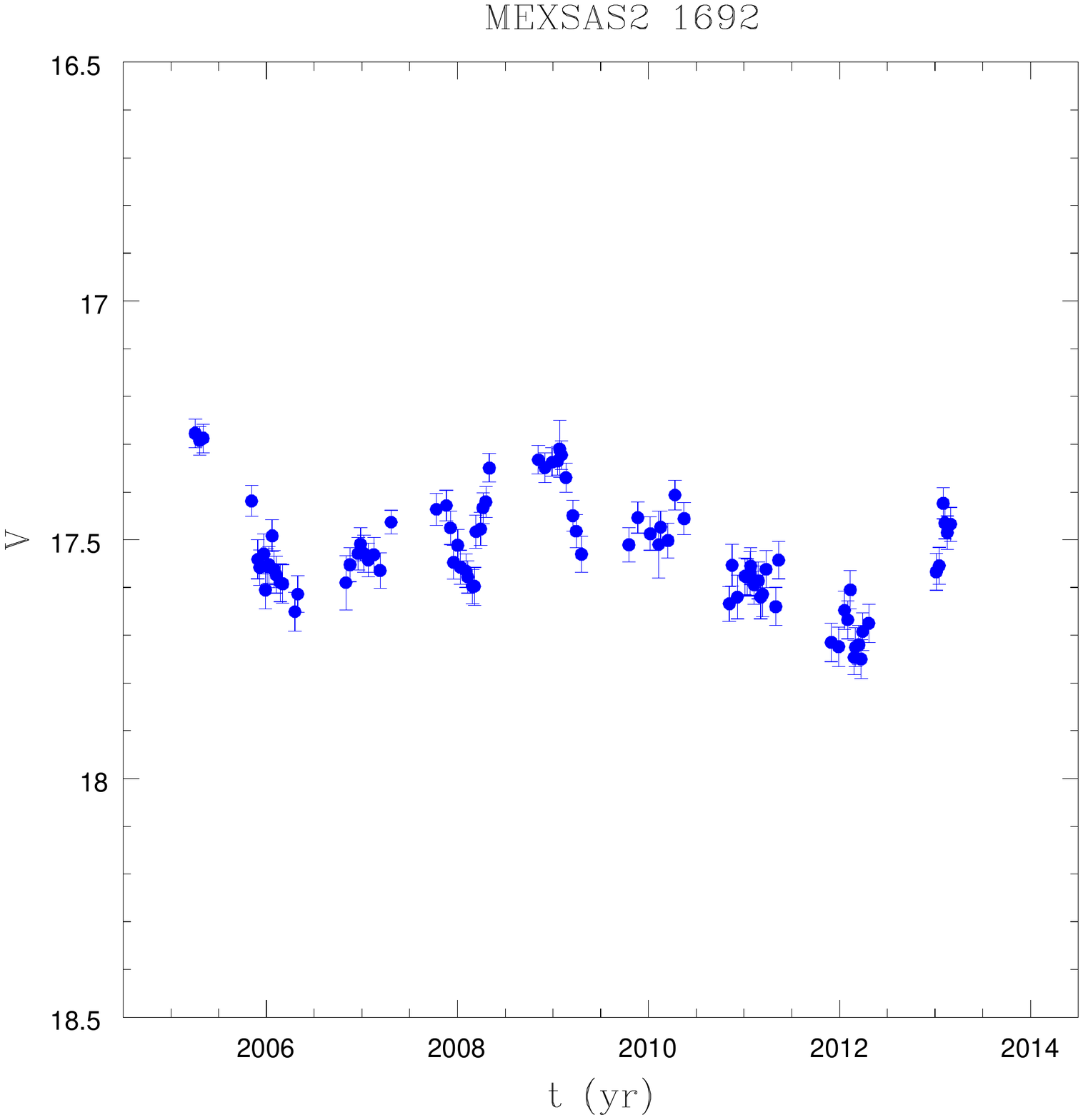}
\vspace{-2cm}
\caption{The optical light curves of MEXSAS2 786 (\emph{left panel}) and MEXSAS2 1692 (\emph{right panel}) computed from their corresponding CSDR2 observations listed in the clean sample.}
\label{fig:lcs}
\end{figure*}

\section{Data}
\label{sec:data}
The present work takes advantage of the Catalina Surveys Data Release 2 (CSDR2)\footnote{\url{http://nesssi.cacr.caltech.edu/DataRelease/}}, which includes multi-epoch observations from both Catalina Sky Survey (CSS) and Catalina Real-Time Transient Survey \citep[CRTS;][]{drake2009}.  
The CSDR2 catalogue lists photometric measurements of $\sim 500$ million objects and hundreds of thousands of extragalactic sources \-- such as AGNs \-- are included among them.
The data are unfiltered, to maximise the amount of incident light, and then calibrated to Johnson's $V$ band \citep{drake2013}.  
CSDR2 is suitable to analyse the variability of individual objects, since each source in this catalogue has been observed several (from $20$ to more than $500$) times through approximately a decade of monitoring activity.

The AGN sample whose optical variability properties we aim to investigate, derives from the crossmatching of CSDR2 with the Multi-Epoch XMM Serendipitous AGNs Sample 2 \citep[MEXSAS2,][]{serafinelli2017}.
The MEXSAS2 catalogue represents the updated version of MEXSAS \citep{vagnetti2016} and contains 9735 detections of 3366 different sources.
This updated sample, MEXSAS2, is constructed from a crossmatching procedure between the sixth data release of the third XMM-Newton serendipitous source catalogue \citep[3XMM-DR6;][]{rosen2016} with two quasar catalogues of the Sloan Digital Sky Survey, that is, SDSS-DR7Q \citep{dr7q} and SDSS-DR12Q \citep{dr12q}. 
MEXSAS2 is also complemented with measurements of characteristic physical quantities of AGNs such as bolometric luminosity ($L\ts{bol}$), Eddington ratio ($\lambda\ts{Edd}$) and black hole mass ($M\ts{BH}$) derived from the catalogues of \cite{shen2011} and \cite{kozlowski2017}, modified by the use of homogeneous criteria, as described in \citet{serafinelli2017}.

The ensemble X-ray variability features of MEXSAS2 have already been extensively studied. Thus, in the present work, we are interested in deriving the individual variability properties of those AGNs composing such catalogue, whose optical counterparts are stored in CSDR2. 
In practice, using the CSDR2 multi-object search service\footnote{\url{http://nesssi.cacr.caltech.edu/cgi-bin/getmulticonedb_release2.cgi}}, we perform a crossmatch between the sources included in MEXSAS2 and CSDR2 within a radius of 3 arcsec in coordinates.
This choice is supported by the following evidence. We repeat the cross-correlation with a set of false coordinates, shifted by 1 arcmin in declination with respect to the true coordinates. Then, we consider the histograms of the mean separation in coordinates for both sets of matches, that is, the one derived from the original coordinates and the other with the artificial ones.
These two histograms cross at approximately 3 arcsec, so we decide to set this as a threshold to minimise possible spurious identifications, while maximising the completeness of the sample. 
This condition is also in agreement with some previous works \citep[e.g.][]{graham2014, suberlak2017}.

Additionally, since MEXSAS2 includes optical photometric data from SDSS, we also impose that the difference between the average magnitude reported by CSDR2 and SDSS\footnote{This is obtained by computing the $V$ magnitude from the $g$ and $r$ SDSS magnitudes, according to \citet{jester2005}.} has to be smaller than 0.8 mag.
This threshold is arbitrary but it takes account of typical variations of few tenths of magnitude over a timescale of years, and of a photometric uncertainty for which the CSDR2 catalogue stores values of the same order.
After the crossmatching procedure, we obtain a catalogue of 340051 observations of 2070 AGNs taken between April 2005 and January 2014, which we will refer to as \emph{matched sample}.
While inspecting the light curves of each source in our sample, it came out that a small fraction of them was occasionally affected by the presence of some outliers, possibly due to spurious detections.
First, we deal with these outliers with a systematic approach. 
Indeed, for each light curve, we look to those detections within few months from the observation which is currently considered. Then we estimate their mean magnitude, $\langle m \rangle$, throughout such time interval, as well as their mean separation in coordinates from the reference values in MEXSAS2, $\langle sep \rangle$.
Ultimately, if the given considered detection lies beyond $3\sigma$ from the previously mentioned $\langle m \rangle$ or if its coordinates are further than $\langle sep \rangle$ from the reference values at $3\sigma$ c.l., such observation is automatically removed from the light curve.
Secondarily, visual inspection has been useful to check that each light curve has been correctly refined.
Furthermore, according to \citet{graham2017}, the published error model for CSDR2 is incorrect: errors at the brightest magnitudes are overestimated and those at fainter magnitudes ($V > 18$) are underestimated \citep[e.g.][]{palaversa2013, vaughan2016}.
This inconsistency between the declared uncertainties and the actual ones is fixed by the implementation of a multiplicative corrective factor as described by \citet{graham2017}. 
In Sect. \ref{sec:sf} we will discuss how the SF is highly sensitive to the effect of the photometric errors.

It has already been pointed out that CSDR2 contains a large amount of detections.
This has been made possible thanks to the intensive monitoring schedule of the Catalina Surveys, which often supplies several observations of the same target in a single day \citep[e.g.][]{mahabal2011}. The drawback is that, sometimes, the photometric uncertainty of a single detection may be quite large (up to $\sim 0.4$ mag). 
It is clear that lowering these single-epoch errors would be desirable. 
We proceed in such direction by binning those points in the light curve of each AGN in the matched sample, which are related to observations collected within a single day.
Thanks to this method, we can study light curves that can rely on a higher level of accuracy.
For instance, many AGNs have been observed four or even more times during the same night.
We are not interested in analysing variability on such short timescales, but the use of binned data allows us in many cases to reduce the photometric uncertainty by a factor $\sim2$.
In any case, it would be difficult to study in detail such short-term variability, since its expected mean value can often be smaller than the photometric errors.
Furthermore, we require that the resulting binned light curves must have a sufficiently large statistics.
For our purposes, this condition is fulfilled whenever the number of epochs when an individual source has been observed, $N\ts{epo}$, is such that $N\ts{epo} \geqslant 40$. 
The new catalogue obtained after the refinement of the light curves and the binning procedure, together with the above mentioned constraint on the statistics of observations, will be labelled as \emph{clean sample}.
The clean sample includes 83304 observations of 1181 AGNs.
As an example, in Fig.\ \ref{fig:lcs} are shown the binned optical light curves of the sources MEXSAS2 786 and MEXSAS2 1692, both included in the clean sample, computed from their corresponding CSDR2 data.

{
\renewcommand{\arraystretch}{1.5}
\begin{table}
\centering
\caption{A comprehensive list of the main steps leading to the creation of the final sample. The elements in the left column represent the catalogues we use in our analysis, sorted by chronological order. The numbers on the right describe the amount of sources included in each sample.}
\label{tab1}
\begin{tabular}{p{1.2cm} p{4.7cm} l } 
 \hline\hline
 Sample & Description & \# Sources \\
 \hline
 Parent sample & MEXSAS2 \citep{serafinelli2017} & 3366 \\
 \addlinespace
 Matched sample & MEXSAS2 sources with optical data from CSDR2 & 2070 \\
 \addlinespace
 Clean sample & $N\ts{epo} \geqslant 40$, removal of spurious identifications & 1181 \\
 \addlinespace
 Reference sample & $N\ts{bins} \geqslant 4$, $\gamma > 0$, available $M\ts{BH}$, $L\ts{bol}$, $\lambda\ts{Edd}$ (see Sect.\ \ref{sec:sf}) & 795 \\
 \hline
\end{tabular}
\end{table}
}

\begin{figure}
\vspace{-1cm}
\centering
\includegraphics[width=\columnwidth]{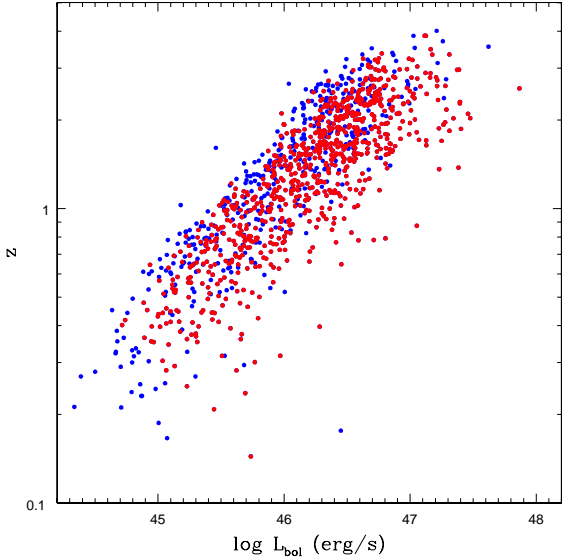}
\vspace{-2.5cm}
\caption{Distribution of the sources in the $L\ts{bol}$$-z$ plane. Blue dots represent the AGNs included in the clean sample. Red dots are related to the objects in the reference sample.}
\label{fig:distrLz}
\end{figure}

\begin{figure}
\vspace{-1cm}
\centering
\includegraphics[width=\columnwidth]{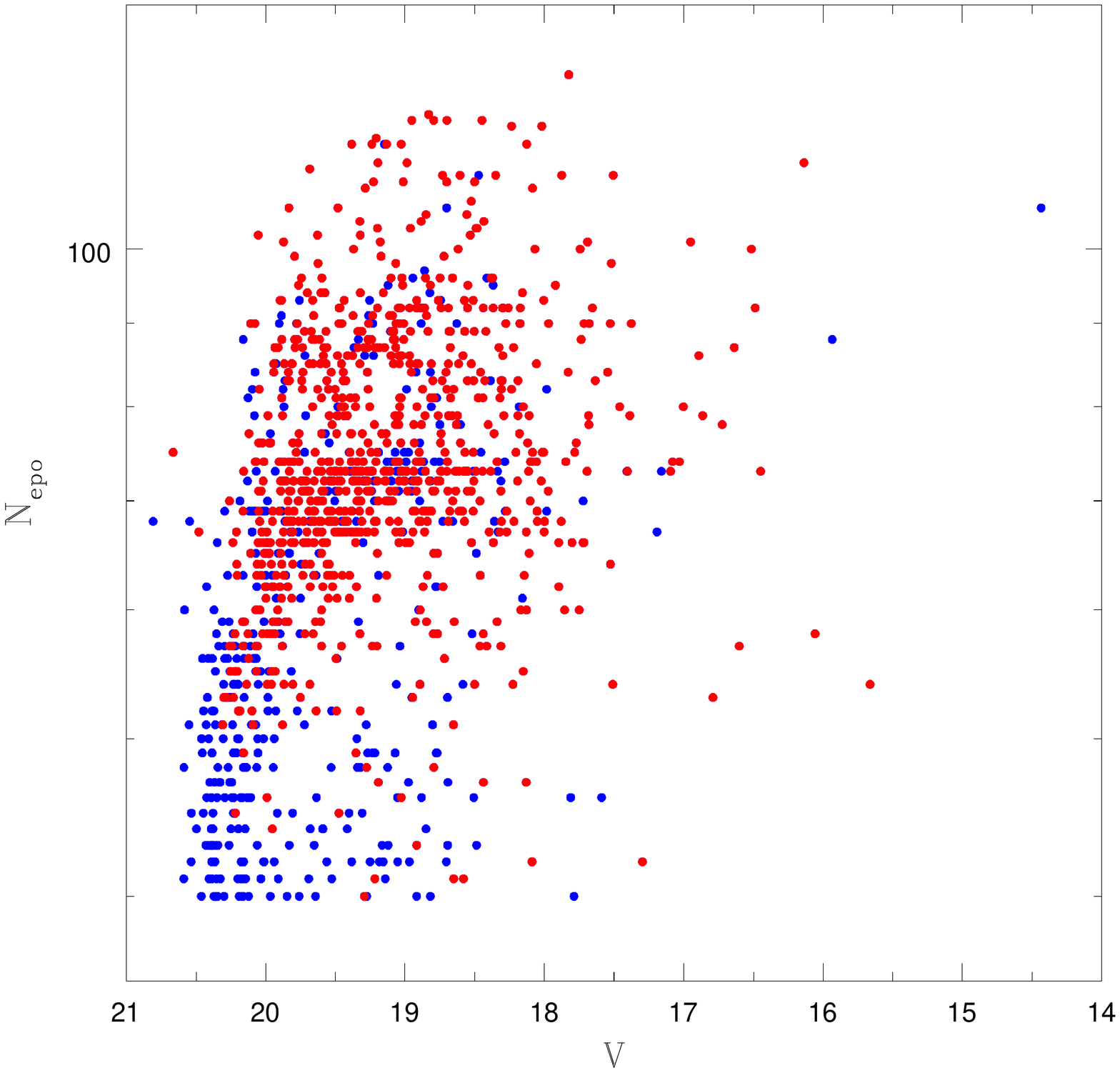}
\vspace{-2.5cm}
\caption{Distribution of the sources in the $V$$-N\ts{epo}$ plane. Blue dots represent the AGNs included in the clean sample. Red dots are related to the objects in the reference sample.}
\label{fig:nepoV}
\end{figure}

The computation of the SF for the objects in the clean sample leads to the ultimate step towards the definition of the \emph{reference sample}, that is, the AGN sample adopted for our analysis.
Such step consists in considering only the most reliable structure functions.
We defer the discussion on this issue to the following section.
In Tab.\ \ref{tab1} it is described a list of the samples used in this work.
Fig.\ \ref{fig:distrLz} shows the distribution of the sources included in the clean sample and in the reference sample in the $L\ts{bol}$$-z$ plane.  
The distribution of the same sources in the $V$$-N\ts{epo}$ plane is shown in Fig.\ \ref{fig:nepoV}.

\section{Structure function}
\label{sec:sf}
\begin{figure*}
\vspace{-1cm}
\centering
\includegraphics[width=\columnwidth]{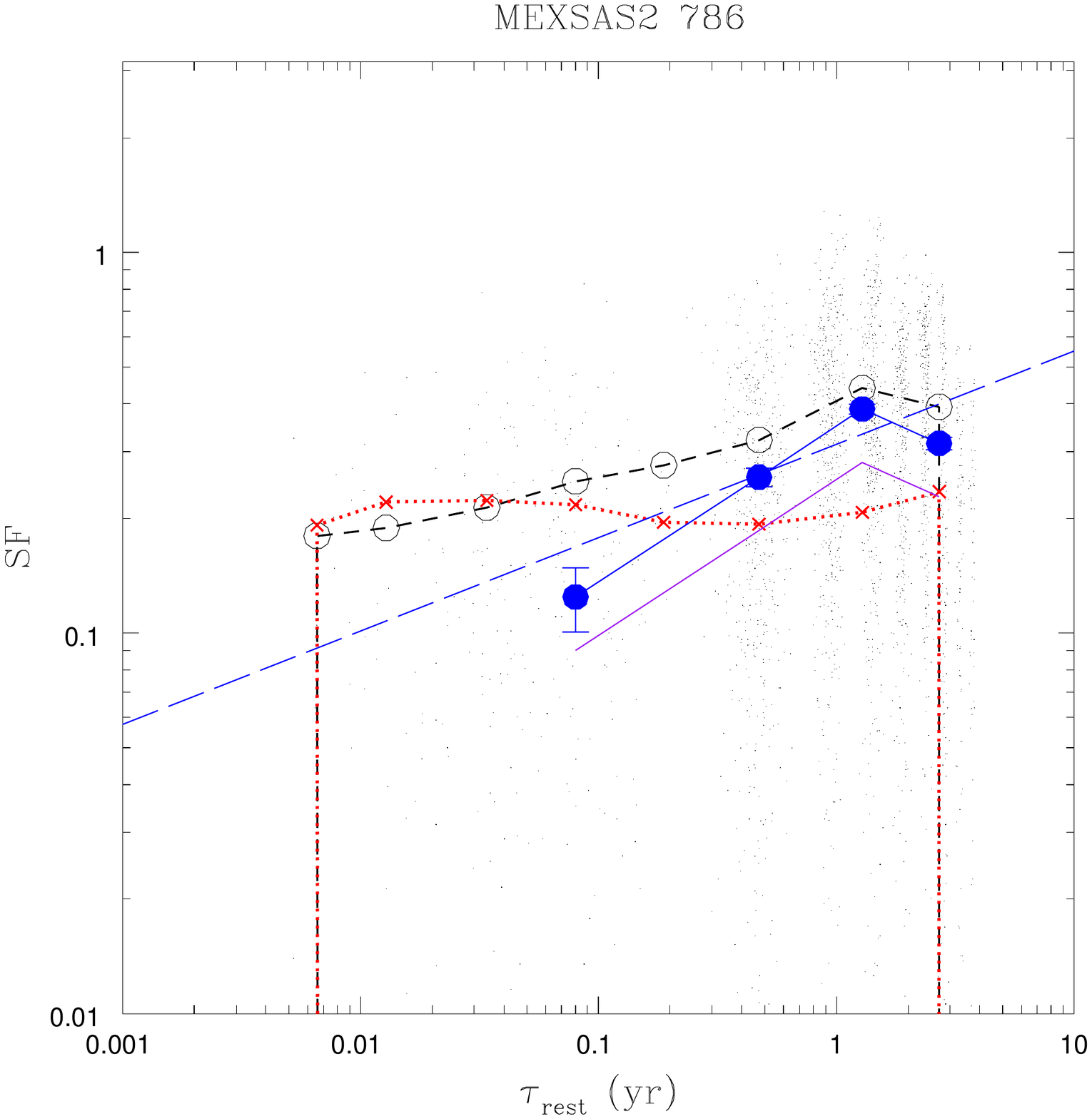}
\includegraphics[width=\columnwidth]{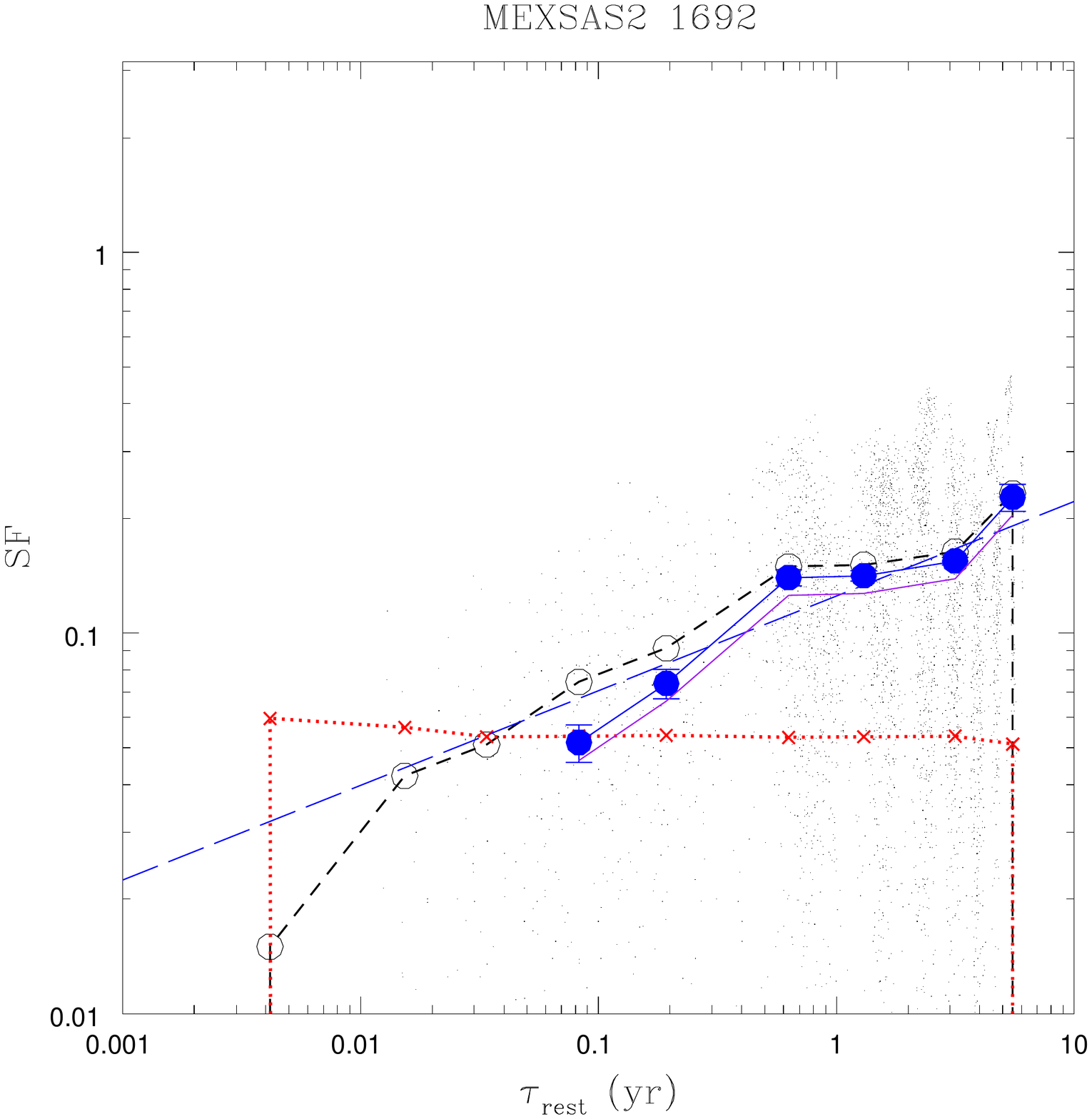}
\vspace{-2cm}
\caption{\emph{Left panel}: SF of MEXSAS2 786. \emph{Right panel}: SF of MEXSAS2 1692. In both panels: empty circles connected by the black short-dashed line show the uncorrected SF. Red crosses connected by the red-dotted line represent the contribution of the photometric errors. The SF corrected for the photometric uncertainties is represented by blue circles connected by the blue solid line. Blue long-dashed line indicate the corresponding least-squares fit to the corrected SF. The fit is weighted for the number of lag values included in each available bin. Purple solid line represents the V-corrected SF (see Sect.\ \ref{sec:Vc}). Small black dots represent the variations for the individual pairs of measurements contributing to the SF. The time lag $\tau$ is evaluated in the rest frame of the given source.}
\label{fig:sfind}
\end{figure*}

The structure function (SF) analysis is a model-independent technique that allows to extract the variability properties of AGNs embedded in their light curves.
The SF works in the time domain and it does not suffer from the windowing and aliasing issues associated to the Fourier techniques \citep[e.g.][]{hughes1992, markowitz2006}.
The standard expression was first introduced by \citet{simonetti1985} in the radio band and, in the optical, it can be written as:
\begin{equation}
    \text{SF}(\tau) \equiv \sqrt{\langle\,[\,m(t + \tau) - m(t)\,]^2 \,\rangle - \sigma^2_{\text{noise}}}\,.
    \label{eq:sf}
\end{equation}

\noindent In the above expression, $m(t)$ and $m(t+\tau)$ are two magnitude measures in a given photometric band, taken at two epochs differing by a time lag $\tau$ in the rest frame of the source.
The term $\sigma^2_{\text{noise}} = \langle [\delta m(t)]^2 + [\delta m(t+\tau)]^2 \rangle$, $\delta m$ being the error on the magnitude at each given time, represents the quadratic contribution of the photometric noise to the observed variations. The photometric noise term needs to be computed carefully as described above, otherwise it may lead to incorrect estimates of the SF \citep[see also the discussion by][]{kozlowski2016}.
The average, indicated by $\langle\dots\rangle$, is always computed within an appropriate bin of time lag around $\tau$.  
The SF is often modelled in terms of a power-law, as:
\begin{equation}
    \text{SF}(\tau) = A\,\bigg(\,\frac{\tau}{\tau_0}\,\bigg)^\gamma \,,
    \label{eq:sfpow}
\end{equation}

\noindent where $A$ and $\gamma$ are two constants \citep[e.g.][]{kawaguchi1998,schmidt2010, peters2015} and, in our work, we choose $\tau_0=1$ yr.
\subsection{Individual SF}
First, we use Eq.\ \ref{eq:sf} to compute the SF of the AGNs included in the clean sample.
Then the SF is fitted by the power-law of Eq.\ \ref{eq:sfpow}, through a linear least-squares fit of the logarithms, weighted for the number of individual lag values falling in each bin.
We remind that the AGNs in the clean sample have been observed $N\ts{epo}$ times, with $N\ts{epo}\geqslant40$.
This means that the number of individual lag values, namely all pairs of epochs contributing to the SF, is $N\ts{pairs}=N\ts{epo}\,(N\ts{epo}-1)/2\geqslant780$.
The whole explored interval of time lags spans from $\sim$1 day to few years. Hence, we divide it in 9 bins, whose width is set to 0.4 dex, so that on average we expect that $N\ts{pairs}/9\gtrsim90$.

Additionally, we require that the SF must be estimated in at least four bins of time lag $\tau$, $N\ts{bins} \geqslant 4$. This condition allows us to focus only on those SFs with the largest statistics, thus with the most reliable fits.
Nevertheless, the contribution of the photometric uncertainties from CSDR2 data is still quite large, especially for the faintest AGNs (see left panel of Fig.\ \ref{fig:sfind}).
This does not prevent the analysis from supplying consistent results which, however, can be less solid than those deriving from the brightest sources (see right panel of Fig.\ \ref{fig:sfind}).  
Indeed, in this case, the photometric errors are lower and it is possible to compute very robust SFs.  

Futhermore, we impose that the slope of the power-law SF must be positive, $\gamma>0$.
This threshold stems from the ubiquitously-reported evidence that variability is an increasing function of the time lag \citep[e.g.][]{hawkins2002,kozlowski2010,sartori2018}, at least until moderate-to-long time lags, where a break is expected, due to the finite size of the emitting region. In the same time, we are quite confident we are not introducing a strong bias, since the sources with unrealistic negative values of $\gamma$ and with a SF satisfying the above condition on $N\ts{bins}$, represent a small fraction of the clean sample ($8 \%$).
Ultimately, we require that each source must be supplied with available meaningful measurements of $L\ts{bol}$, $\lambda\ts{Edd}$, $M\ts{BH}$. 
Once the above conditions are applied to the clean sample, this leads to the definition of the \emph{reference sample}, which includes 59976 observations of 795 AGNs. 

\begin{figure}
\vspace{-1cm}
\centering
\includegraphics[width=\columnwidth]{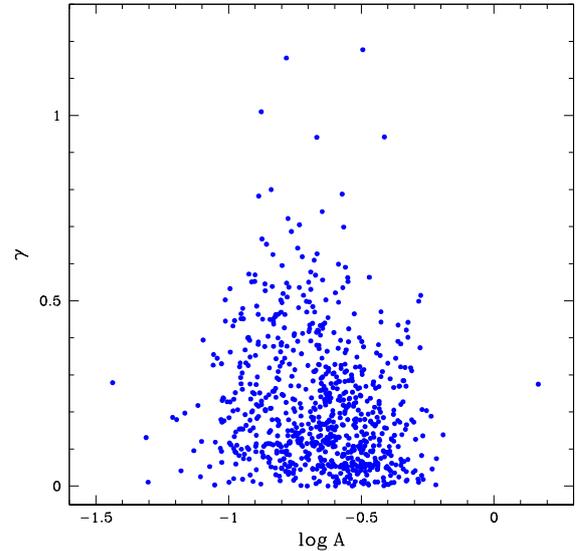}
\vspace{-2.5cm}
\caption{Distribution of the variability parameters returned by the least-squares fit of the power-law SF for each source in the reference sample.
These parameters are both computed from the time lags in the rest frame of each source.}
\label{fig:ab}
\end{figure}

In Fig.\ \ref{fig:ab} it is shown the distribution of the individual variability parameters in the reference sample.
It is important to underline that $\log{A}$ represents the variability amplitude of a given source taken at a rest-frame time lag $\tau=1$ yr, while $\gamma$ describes and quantifies the way in which variability amplitude grows with increasing time lag.

\noindent A distribution similar to that shown in Fig.\ \ref{fig:ab} can also be found in the works of \citet{schmidt2010} and \citet{peters2015}. The primary goal of their investigations, however, was not oriented towards an individual variability study of intrinsic properties of AGNs.
Indeed, they only used the SF analysis to identify potential AGN candidates thanks to their observed variability properties.
In fact, computing the SF, they considered the time lags in the observer frame. This slightly modifies the distribution of Fig.\ \ref{fig:ab}, since $\tau\ts{obs} = \tau\,(1+z)$ with $\tau$ being the rest-frame time lag.
From Eq.\ \ref{eq:sfpow} it follows that:
\begin{equation}
    \text{SF}(\tau\ts{obs}) = A\ts{obs}\,\bigg(\frac{\tau\ts{obs}}{\tau_0}\bigg)^\gamma\,,\quad A\ts{obs} = A\,(1+z)^{-\gamma}\,.
    \label{eq:tobs}
\end{equation}

\begin{figure}
\vspace{-1cm}
\centering
\includegraphics[width=\columnwidth]{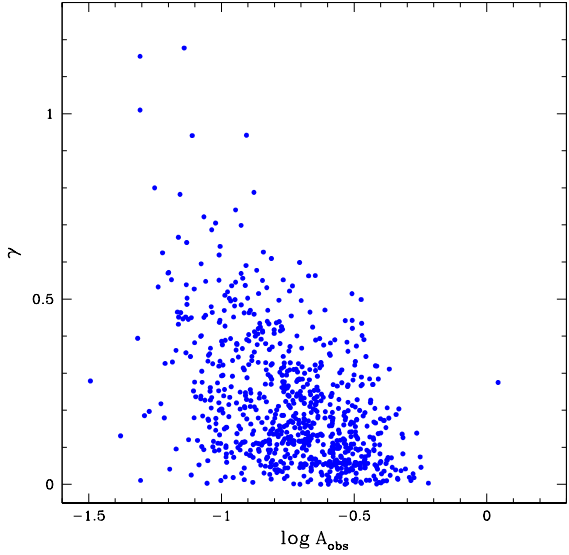}
\vspace{-2.5cm}
\caption{Distribution of the variability parameters of the AGNs in the reference sample, computed by considering the observer-frame time lags.} 
\label{fig:ab_obs}
\end{figure}

\noindent This means that when we move from the rest frame to the observer frame, the variability amplitude parameter is modified according to Eq.\ \ref{eq:tobs}, while the slope of the SF, $\gamma$, is clearly not affected.
In Fig.\ \ref{fig:ab_obs}, as a mere comparison, it is shown the result of such transformation, displaying a distribution which is very similar to those reported in \citet{schmidt2010} as well as in \citet{peters2015}.
Nevertheless, in the context of a variability analysis, it would be preferable to consider the rest-frame properties of the given source. In other words, the representation of the distribution of the variability parameters in Fig.\ \ref{fig:ab} is to be preferred to that of Fig.\ \ref{fig:ab_obs}, since the former is derived from considering the rest-frame time lags for each AGN.

\subsection{V-correction}
\label{sec:Vc}
We have already discussed in Sect.\ \ref{sec:intro} that variability appears to be frequency-dependent. More specifically, in the optical band it has been ubiquitously reported a bluer-when-brighter trend, that is, larger flux variations are observed at bluer rest-frame frequencies \citep[e.g.][]{cai2016}.
At present, several indicators have been proposed to quantify this spectral variability effect \citep[e.g.][]{trevese2002, kokubo2014, sun2014}. In particular, \citet{trevese2002} introduced the $\beta$ parameter as:

\begin{equation}
    \beta = \frac{\Delta \alpha}{\Delta{\log{f_\nu}}}\, ,
\end{equation}

\noindent where $f_\nu$ is the monochromatic flux in a given band and $\alpha$ is the spectral index, considering $f_\nu \propto \nu^{\alpha}$.
The above parameter quantifies the variations of the spectral index in correspondence of different flux variations. A value of $\beta>0$ indicates the previously mentioned bluer-when-brighter trend.
Furthermore, $\beta$ is intimately connected to the dependence of the SF on the rest-frame frequency, as discussed (for the X-ray band) by \citet{vagnetti2016}.

Indeed, they found that:
\begin{equation}
    \beta = (\log{e})^{-1}\cdot\delta{\log{\text{SF}}}/\delta{\log{\nu}}\,.
    \label{eq:beta}
\end{equation}

\noindent In this way, \citet{vagnetti2016} calculated such spectral variability estimator in their multi-wavelength ensemble X-ray variability analysis.
Turning back to the optical, the CSDR2 photometry is unfortunately limited to a single band and we do not have the possibility to compute independently the value of $\beta$.
It should be noticed that, in principle, this value can be different from one source to another (see \citet{trevese2002}).
However, we rely on the estimate extrapolated from the optical AGN variability analysis of \citet{morganson2014}, where they found that $\text{SF}\propto\lambda^{-0.441}$. Once the above relation is converted in terms of $\beta$ we conclude that, on average, an acceptable compromise is to set $\beta=1$ for all AGNs in the reference sample.   
This is also in agreement with the results of several other similar studies found in the literature \citep[e.g.][]{diclemente1996,schmidt2012, sun2014}.

The above discussion was necessary since the dependence of variability on frequency implies that the corresponding flux variations in the rest frame are not the same as those estimated in the observer frame.
In order to clarify this issue, let us focus on Fig.\ \ref{fig:ab}. In this plot we have converted the time lags in the rest frame of the source while leaving unchanged the estimation of variability.
However, our analysis of flux variations is based on data collected in observer-frame bands.
It is not possible to fix the rest-frame band for AGNs at different redshifts but this limitation can be overcome by considering the estimated spectral variability (Eq.\ \ref{eq:beta}).
This allows us to simulate the shift from the observer frame to the rest frame.
For a source at a given redshift $z$, we measure variability in a rest-frame band shifted by $\delta{\log{\nu}}=\log{(1+z)}$.
Thus, from Eq.\ \ref{eq:beta}, it follows that:
\begin{equation}
    \delta{\log{\text{SF}}}  \simeq \log{e}\cdot\beta\log{(1+z)}\,.
\end{equation}

\noindent The effect consists in an overall upwards shift (since $\beta>0$) of the individual SF of each source included in a given sample. 
Hence, it is necessary to apply an opposite downwards correction, which \citet{vagnetti2016} introduced for the first time with the name of V-correction, where V stands for variability:
\begin{equation}
    \text{V-corr} \equiv -\delta{\log{\text{SF}}} \simeq -\log{e}\cdot\beta\log{(1+z)}\,.
    \label{eq:Vcorr7}
\end{equation}

\noindent It is important to underline that the standard K-correction has no effect on the computation of the SFs. Indeed all measured fluxes, before and after a variation, are affected by the same $z$-dependent factor for any given source. It follows that the corresponding logarithmic change is not altered.

Since in our work we adopt $\beta=1$, the V-corrected variability amplitude parameter can be expressed as:
\begin{equation}
    \log{A\ts{V-corr}} \simeq \log{A} - 0.434\cdot\log{(1+z)}\,.
    \label{eq: vcorr}
\end{equation}

\begin{figure}
\vspace{-1cm}
\centering
\includegraphics[width=\columnwidth]{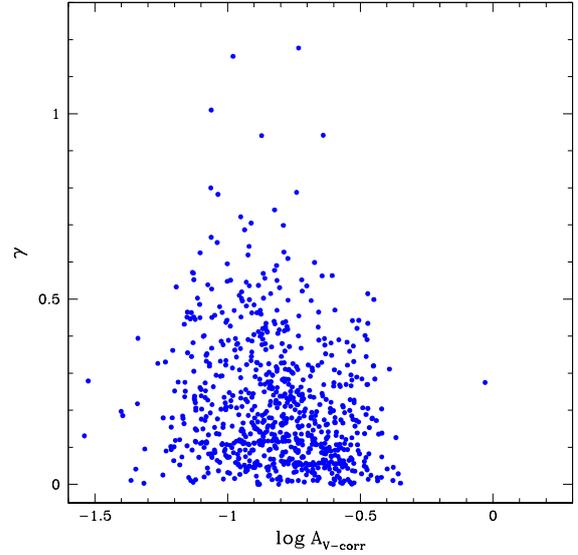}
\vspace{-2.5cm}
\caption{Distribution of the variability parameters of the AGNs in the reference sample after the V-correction.} 
\label{fig:ab_Vcorr}
\end{figure}

\noindent Clearly, this does not affect the slope parameter, $\gamma$, which remains unchanged.
In Fig.\ \ref{fig:ab_Vcorr} it is shown the overall effect of the V-correction on the distribution of variability parameters of the AGNs in the reference sample.
From a simple comparison with Fig.\ \ref{fig:ab}, we can now notice a global shift towards smaller values of variability amplitude. This effect can also be appreciated in terms of the V-corrected individual SFs shown in Fig.\ \ref{fig:sfind}, where the effect is stronger for MEXSAS2 786 ($z=1.1$) compared to MEXSAS2 1692 ($z=0.28$).

{
\begin{table*}
\LARGE
\centering
\caption{Overall description of the reference sample.\\}
\label{tab:cds}
\renewcommand{\arraystretch}{1.5}
\begin{adjustbox}{max width=\textwidth}
\begin{threeparttable}
\begin{tabular}{M{1cm} c M{1.5cm} M{1.5cm} M{1.5cm} M{2.5cm} M{1.8cm} M{2.0cm} M{1.0cm} c M{1.5cm} c M{2.0cm} M{2.0cm} M{1.2cm} M{1.8cm} M{1.8cm} c c}
\hline\hline
ID &  SDSSJ & Ra (deg) & Dec (deg) &  $\log{A}$ &  $\log{A\ts{V-corr}}$ & $\sigma_{\log{A\ts{V-corr}}}$ & $\gamma$ & $\sigma_\gamma$ & $R^2$ & $z$ & $\log{(M\ts{BH}/M_\odot)}$ & $\log{L\ts{bol}}$ (erg/s)& $\log{\lambda\ts{Edd}}$ & $V$ & $\log{L\ts{V}}$ (erg/s) &$\log{L\ts{X}}$ (erg/s) & $N\ts{epo}$ & $N\ts{epx}$ \\
\hline

 3 &  000557.64+195749.2   &  1.490	&	19.964	&	-0.88	&	-1.13	&	0.02	&	0.01	&	0.05	& 0.004 &	2.78	&	9.30	&	46.88	&	-0.52	&	18.91	&	45.97	&	44.73	&	72	&	2  \\
 
 15     &      003027.76+261356.5      &        7.616	&	26.232	&	-0.77	&	-1.04	&	0.07	&	0.1	&	0.2	& 0.074 &	3.24	&	9.34	&	46.75	&	-0.68	&	19.87	&	45.69	&	44.49	&	74	&	3    \\ 
 
  18       &    003109.45+261004.7      &        7.789	&	26.168	&	-0.33	&	-0.45	&	0.02	&	0.32	&	0.06	& 0.905 &	0.84	&	9.07	&	45.66	&	-1.51	&	19.83	&	44.63	&	44.51	&	68	&	3   \\  
  
  24     &      004243.06+000201.3     &  10.679	&	0.034	&	-1.11	&	-1.24	&	0.04	&	0.03	&	0.09	& 0.017 &	1.08	&	8.84	&	46.48	&	-0.46	&	18.00	&	45.58	&	44.29	&	67	&	2   \\  
  
  27       &    004316.86+001008.4     &         10.820	&	0.169	&	-0.91	&	-1.08	&	0.06	&	0.3	&	0.1	& 0.508 &	1.52	&	9.08	&	46.56	&	-0.64	&	18.12	&	45.81	&	44.60	&	76	&	2  \\   
  
  28     &      004336.00+010636.9    &   10.900	&	1.110	&	-0.76	&	-0.89	&	0.03	&	0.15	&	0.08	& 0.554 &	1.00	&	8.57	&	45.90	&	-0.78	&	19.39	&	44.96	&	44.41	&	75	&	2   \\  

\hline
\end{tabular}

\begin{tablenotes}
      \item Note: the full version of the table is available at CDS.
\end{tablenotes}
\end{threeparttable}
\end{adjustbox}

\vspace{0.5cm}
\end{table*}
}

The variability parameters of the AGNs in the reference sample, along with several other useful information about the sources, are reported in Tab.\ \ref{tab:cds}, where Col.\ 1 indicates the MEXSAS2 serial number; Col.\ 2 the SDSS identifier; Col.\ 3 the right ascension; Col.\ 4 the declination; Col.\ 5 the standard variability amplitude parameter $\log{A}$; Col.\ 6 the V-corrected variability amplitude parameter $\log{A\ts{V-corr}}$; Col.\ 7 the uncertainty on the V-corrected variability amplitude parameter, which is equal to that of the standard parameter; Col.\ 8 the slope parameter of the power-law SF $\gamma$; Col.\ 9 the uncertainty on $\gamma$; Col.\ 10 the coefficient of determination $R^2$, as a measure of the goodness of the fit; Col.\ 11 the redshift; Col.\ 12 the black hole mass; Col.\ 13 the bolometric luminosity; Col.\ 14 the logarithmic Eddington ratio $\log{\lambda\ts{Edd}}$; Col.\ 15 the average $V$ magnitude; Col.\ 16 the $V$-band luminosity; Col.\ 17 the X-ray luminosity in the $0.5-4.5$ keV band; Col.\ 18 the number of individual optical observations in CSDR2 $N\ts{epo}$; Col.\ 19 the number of individual X-ray observations in MEXSAS2 $N\ts{epx}$.  

\subsection{Ensemble SF}
\label{subsec:sfens}

In the previous section we carried out a study of the variability properties connected to individual sources, by means of the SF method. We want to stress once again that a similar approach is allowed by the large amount of data collected by the Catalina Surveys for each observed AGN.
However, this condition is rarely fulfilled by the vast majority of the surveys which are currently operating.
Nevertheless, if one is interested in the variability analysis of a given sample, the SF method is still a viable option.
Indeed, it is possible to compute an ensemble SF by jointly considering all the observations of each source in the given sample.

\begin{figure}
   \vspace{-1cm}
   \centering
   \includegraphics[width=\columnwidth]{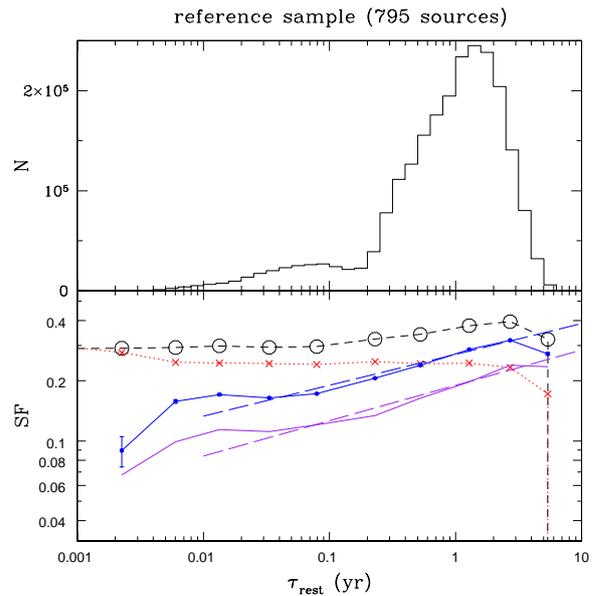}
   \vspace{-2.5cm}
   \caption{Ensemble SF for the reference sample. Purple solid line represents the V-corrected SF. Purple long-dashed line indicates the corresponding least-squares fit to the V-corrected SF. The fit is weighted for the number of lag values included in each available bin. The remaining symbols and lines have the same meaning described for the individual SFs in Fig.\ \ref{fig:sfind}. The time lag $\tau$ is evaluated in the rest frame for each given source. The histogram on top shows the distribution of the individual pairs of epochs contributing to the SF as a function of the time lag.}
   \label{fig:sfens}
\end{figure}

In order to provide a full description of the variability properties of our AGN sample, we compute the ensemble SF for the reference sample (see Fig.\ \ref{fig:sfens}).
The least-squares fit returns a value of $\gamma = 0.15\pm0.01$ and a variability amplitude equal to $\log{A}=-0.568\pm0.006$.
These outcomes are compatible with other similar works in the literature \citep[e.g.][]{devries2005, li2018}.

However, in the previous section we called attention to the importance of estimating both variability and time lags in the rest frame of the sources. This is done thanks to the V-correction. In Fig.\ \ref{fig:sfens} we can appreciate the overall effect of the V-correction on the ensemble SF. 
We underline that, contrarily to what we have seen for the individual SFs, in this case also the slope parameter is affected. Indeed, the slope of the V-corrected SF is slightly steeper than that resulting from the standard computation. 
The reason is that from Eq.\ \ref{eq: vcorr} we can see how higher-$z$ AGNs are affected by a larger correction.
Moreover, AGNs at higher redshift have also smaller rest-frame time lags.
Once combined, these effects tend to steepen the resulting SF by lowering the variability amplitude at small time lags.

\noindent In Tab.\ \ref{tab:sfens} we show the results of the least-squares fit to the standard and V-corrected ensemble SFs for the reference sample.

{
\renewcommand{\arraystretch}{1.5}
\begin{table}
    \centering
    \caption{Results of the least-squares fit to the ensemble SF for the reference sample. The outcomes are distinguished in terms of the methodology behind the computation of the SF: (i) the standard approach; (ii) the V-correction procedure.}
    \begin{tabularx}{\columnwidth}{ c c c c c}
        \hline\hline
       Sample  & Method & $\log{A}$ & $\gamma$ & $R^2$\\
       \hline
       Reference & Standard &  $-0.568\pm0.006$   & $0.15\pm0.01$   &  0.867 \\
       Reference & V-correction &  $-0.722\pm0.008$   &  $0.18\pm0.02$ & 0.946  \\
       \hline
   \end{tabularx}
   \label{tab:sfens}
\end{table}
}

\section{Correlation with physical quantities}
\label{sec:corr}
In Sect.\ \ref{sec:data} we have anticipated that MEXSAS2 was complemented with measurements of physical properties of AGNs. In this section, we aim to search for possible correlations between these quantities and the variability features of the sources in the reference sample.
This is done by considering the two individual variability parameters derived by the SF analysis described in Sect.\ \ref{sec:sf}.
In order to perform a consistent analysis of the rest-frame properties of our sample, we select the V-corrected variability amplitude, $\log{A\ts{V-corr}}$, as the first parameter (see discussion in Sect.\ \ref{sec:Vc}).
The second parameter, $\gamma$, represents the slope of the power-law SF.
In the following, we will discuss the results of several correlations with respect to such parameters. All these correlations are derived by means of a random sampling procedure, that
consists in synthesising a large number ($N=10000$) of random samples from the observed data distribution, while simultaneously accounting for the error distribution of both the dependent and independent variables. In this way, such method combines its simplicity with its capability of providing a robust fit to the given relation. The only drawback, in this case, is represented by the uncertainties that affect physical quantities, such as $M\ts{BH}$ and $L\ts{bol}$. Black hole mass measurements are known to be influenced by very large errors, and there is a general consensus in considering these errors can be up to $\sim 0.5$ dex \citep[e.g.][]{shen2013, rakshit2020}. For this reason, we set uniform uncertainties of $0.5$ dex on the adopted $M\ts{BH}$ estimates. The errors one should consider on the bolometric luminosity are also somewhat arbitrary. Indeed, in principle, we should take into account the uncertainty on both the monochromatic luminosities that are used to derive $L\ts{bol}$ and the corresponding necessary bolometric corrections $k\ts{bol}$. While monochromatic luminosities are usually tabulated with errors around $\sim0.1$ dex, the uncertainty on $k\ts{bol}$ is less clear. According to \citet{richards2006}, objects can have their bolometric luminosity misestimated by up to $50\%$. Assuming a base error of $\sim 0.1$ dex, and allowing for an additional $50\%$ uncertainty on that, it translates into a typical error of $\sim 0.2$ dex, which is thus the one we set on $L\ts{bol}$.
Concerning the redshift $z$, we take its reference value for each source, neglecting its uncertainty, since it is usually determined with high accuracy.

\begin{figure}
\centering
\includegraphics[width=\columnwidth]{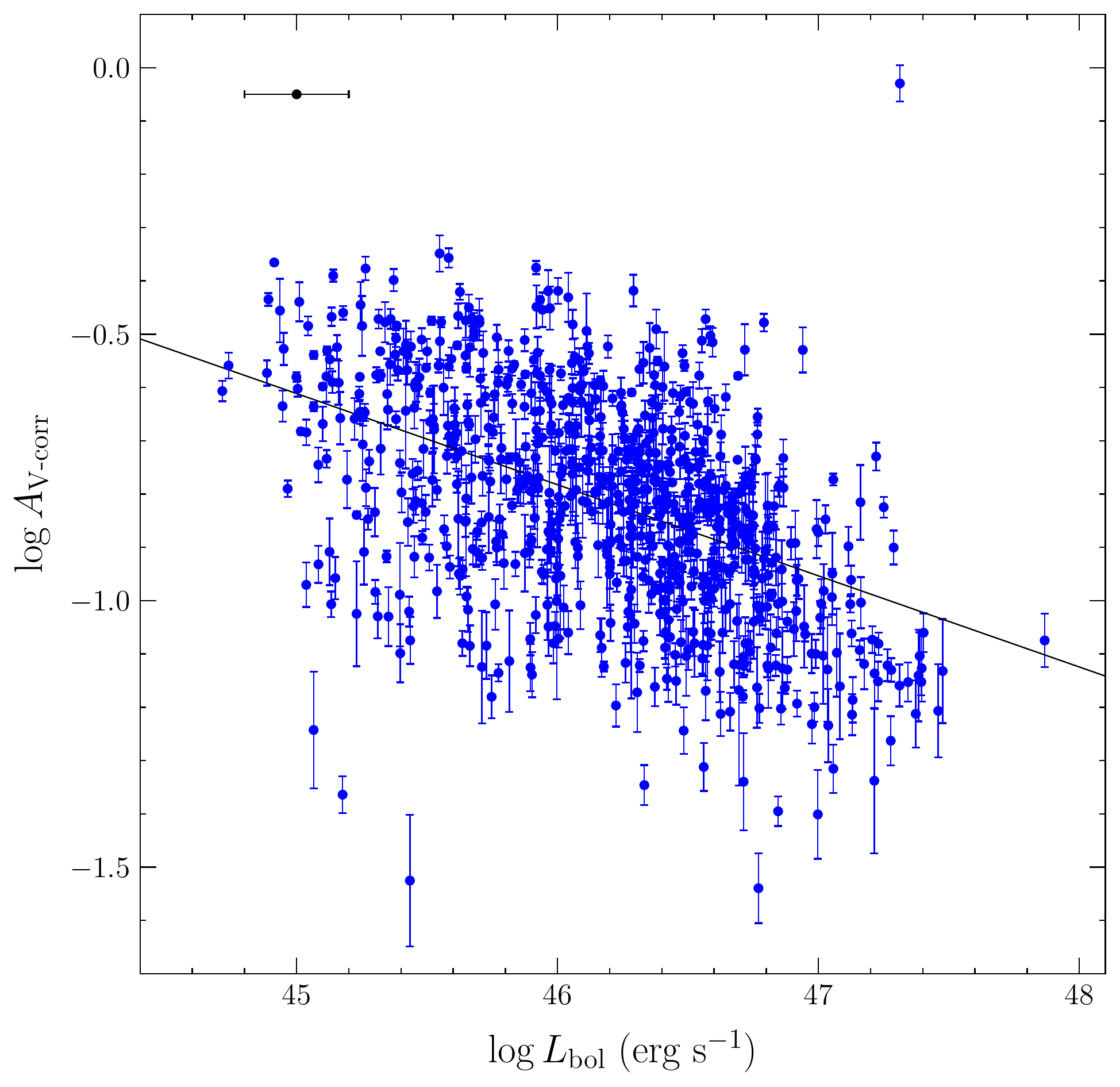}
\caption{Variability amplitude is negatively correlated with the bolometric luminosity. 
The typical error bar of $\log{L\ts{bol}}$ is indicated in black in the upper-left corner. The Spearman's correlation coefficient is $\rho\ts{s}=-0.48$, with null probability $p(>\!|\rho\ts{s}|) < 10^{-10}$. The black solid line describes the best-fit relation.}
\label{fig:lbol_logAVc}
\end{figure}

We find variability amplitude in the rest frame to be negatively correlated with $L\ts{bol}$, as shown in Fig.\ \ref{fig:lbol_logAVc}. The Spearman's coefficient is $\rho\ts{s}=-0.48$, with null probability $p(>\!|\rho\ts{s}|) < 10^{-10}$.  

\noindent The linear least-squares fit returns the following result:
\begin{equation}
    \log{A\ts{V-corr}} = (7.1\pm0.5) - (0.17\pm0.01)\, \log{L\ts{bol}}\,.
    \label{eq:lbol}
\end{equation}

\noindent This is indicating that the most luminous quasars are also the least variable, a result which is in agreement with several studies in the literature \citep[e.g.][]{hook1994, cristiani1996, bauer2009, morganson2014}.

\begin{figure}
\centering
\includegraphics[width=\columnwidth]{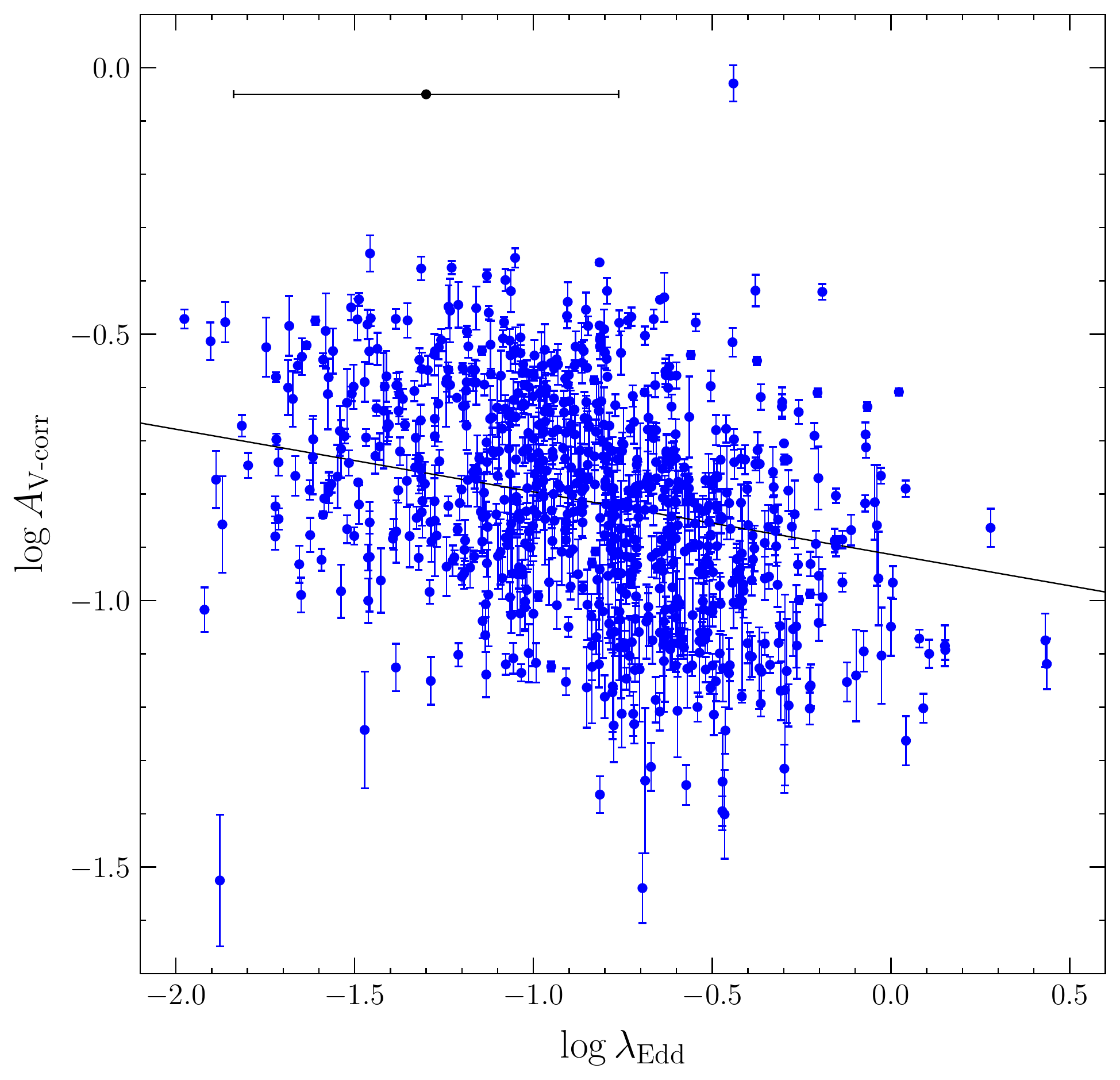}
\caption{Variability amplitude is anti-correlated with the Eddington ratio. 
The typical error bar of $\log\lambda\ts{Edd}$ is indicated in black in the upper-left corner. The Spearman's correlation coefficient is $\rho\ts{s}=-0.30$, with null probability $p(>\!|\rho\ts{s}|) < 10^{-10}$. The black solid line describes the best-fit relation.}
\label{fig:edd_logAvc}
\end{figure}

\noindent From a simple univariate analysis, we also find that variability amplitude is negatively correlated with the Eddington ratio, $\lambda\ts{Edd}$ and the black hole mass, $M\ts{BH}$.
The Spearman's correlation coefficient is $\rho\ts{s}=-0.30$ for the former, with a null probability $p(>\!|\rho\ts{s}|) < 10^{-10}$ and $\rho\ts{s}=-0.21$ for the latter, with a null probability $p(>\!|\rho\ts{s}|) \simeq 5\times10^{-9}$.
We obtain, respectively, that $\log{A\ts{V-corr}}=(-0.91\pm0.01) - (0.12\pm0.01)\,\log{\lambda\ts{Edd}}$ and $\log{A\ts{V-corr}}=(-0.2\pm0.1) - (0.07\pm0.01)\,\log{(M\ts{BH}/M_\odot)}$. 
As an example, in Fig.\ \ref{fig:edd_logAvc} we show the distribution of the variability amplitude parameter with respect to the Eddington ratio.

However, we know that $L\ts{bol}$, $\lambda\ts{Edd}$ and $M\ts{BH}$ are intimately and strictly connected to each other.
This implies that a bivariate analysis is mandatory, in order to disentangle the various individual dependencies related to variability amplitude.
We choose bolometric luminosity and black hole mass as independent variables and we consider the following relation:
\begin{equation*}
    \log{A\ts{V-corr}} = k_1 + k_2\,\text{log}\,\mathcal{L}\ts{bol} + k_3\,\text{log}\,(\mathcal{M}\ts{BH}\,/\, M\ts{$\odot$})\,.
\end{equation*}

\begin{figure}
   \centering
   \includegraphics[width=\columnwidth]{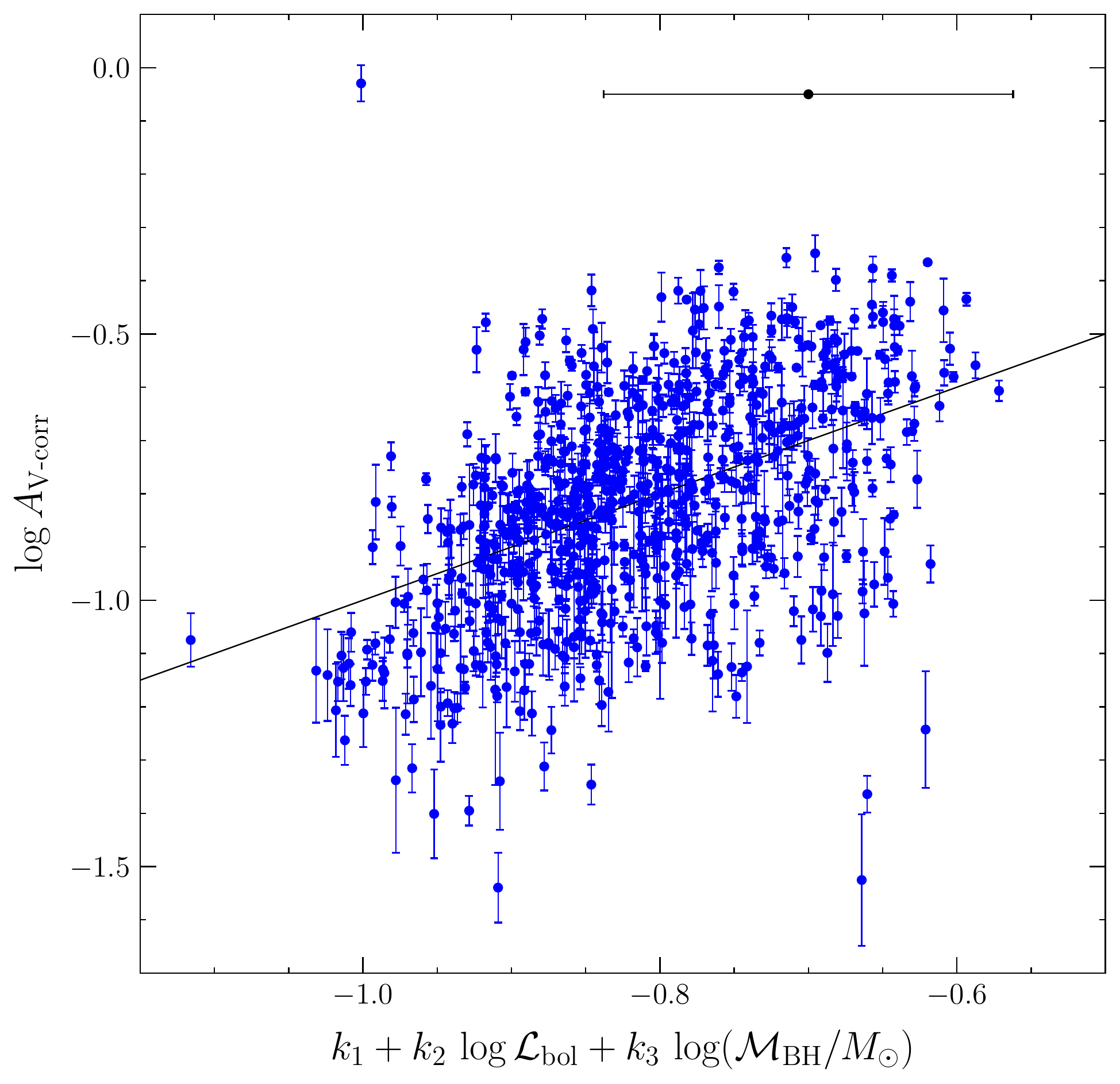}
   \caption{Representation of the variability amplitude parameter, $\log{A\ts{V-corr}}$, vs. its best-fit bivariate relation in terms of the normalised bolometric luminosity and black hole mass. The typical error bar of the bivariate function is indicated in black in the upper-right corner. The $k$ coefficients, which are computed from the best-fit relation described by the black solid line, are those provided in the main text. The Spearman's rank correlation coefficient is $\rho\ts{s} = 0.48$ with null probability $<10^{-10}$.}
   \label{fig:multi_logAvc}
\end{figure}

\noindent We underline that, in the expression above, the calligraphic notation indicates that both bolometric luminosity and black hole mass are normalised with respect to their characteristic range of values. In this way, the different widths of such intervals do not affect the $k$ coefficients, which, in turn, are dimensionless. 
We obtain that $k_1 = -0.61\pm0.02$, $k_2 = -0.59 \pm 0.04$ and $k_3 = 0.11 \pm 0.04$. 
This means that we have a strong indication for the variability amplitude to be anti-correlated with the bolometric luminosity, while we can safely neglect its dependence on $M\ts{BH}$, which is weaker than that related to $L\ts{bol}$. This is also suggesting that the previously reported inverse correlation of $\log{A\ts{V-corr}}$ with $\lambda\ts{Edd}$ may be spurious, since it is likely to be mainly driven by the negative correlation with $L\ts{bol}$. In Fig.\ \ref{fig:multi_logAvc} we show the best-fit bivariate function vs.\ $\log{A\ts{V-corr}}$. The Spearman's rank correlation coefficient for the best-fit bivariate correlation is $\rho\ts{s} = 0.48$, with null probability $p(>\!|\rho\ts{s}|) < 10^{-10}$.

Regarding the slope parameter of the SF, $\gamma$, from the univariate analysis we find no significant correlation with the Eddington ratio. On the contrary, $\gamma$ is positively correlated with both bolometric luminosity ($\rho\ts{s}=0.20$) and black hole mass ($\rho\ts{s}=0.19$), with corresponding null probabilities $p(>\!|\rho\ts{s}|) \simeq 2 \times 10^{-8}$ and $p(>\!|\rho\ts{s}|) \simeq 5 \times 10^{-8}$.
Nevertheless, these trends are very weak. Indeed, we obtain respectively $\gamma = (-3.1\pm0.5) + (0.07 \pm 0.01)\,\log{L\ts{bol}}$ and $\gamma = (-0.3\pm0.1) + (0.06\pm0.01)\,\log{(M\ts{BH}/M_\odot)}$.
In order to determine which of these relations is more important than the other, we carry on a bivariate correlation using the same approach described above. We consider the following expression:
\begin{equation*}
    \gamma = c_1 + c_2\,\text{log}\,\mathcal{L}\ts{bol} + c_3\,\text{log}\,(\mathcal{M}\ts{BH}\,/\, M\ts{$\odot$})\,.
\end{equation*}

\noindent Once again, both bolometric luminosity and black hole mass are properly normalised, so that the resulting coefficients are dimensionless. We obtain that $c_1 = 0.08 \pm 0.02$, $c_2 = 0.17 \pm 0.04$ and $c_3 = 0.08 \pm 0.04$. These outcomes are suggesting that $\gamma$ is mainly correlated with the bolometric luminosity, while its dependence on $M\ts{BH}$ is weaker, though not negligible. The Spearman's rank correlation coefficient for the best-fit bivariate correlation is $\rho\ts{s} = 0.21$, with null probability $p(>\!|\rho\ts{s}|) \simeq 10^{-9}$.

With reference to the relation between variability amplitude and redshift, we must remember that in every flux-limited sample, $z$ is strongly correlated with luminosity.
This is why it is necessary to disentangle the effects of the above dependencies.
A viable approach consists in considering the residuals of the $\log{A\ts{V-corr}}$$-\log{L\ts{bol}}$ relation, defined as:
\begin{equation}
    \Delta{\log{A\ts{V-corr}}} = \log{A\ts{V-corr}} - \log{A\ts{V-corr}}(L\ts{bol})\,.
\end{equation}

\begin{figure}
\centering
\includegraphics[width=\columnwidth]{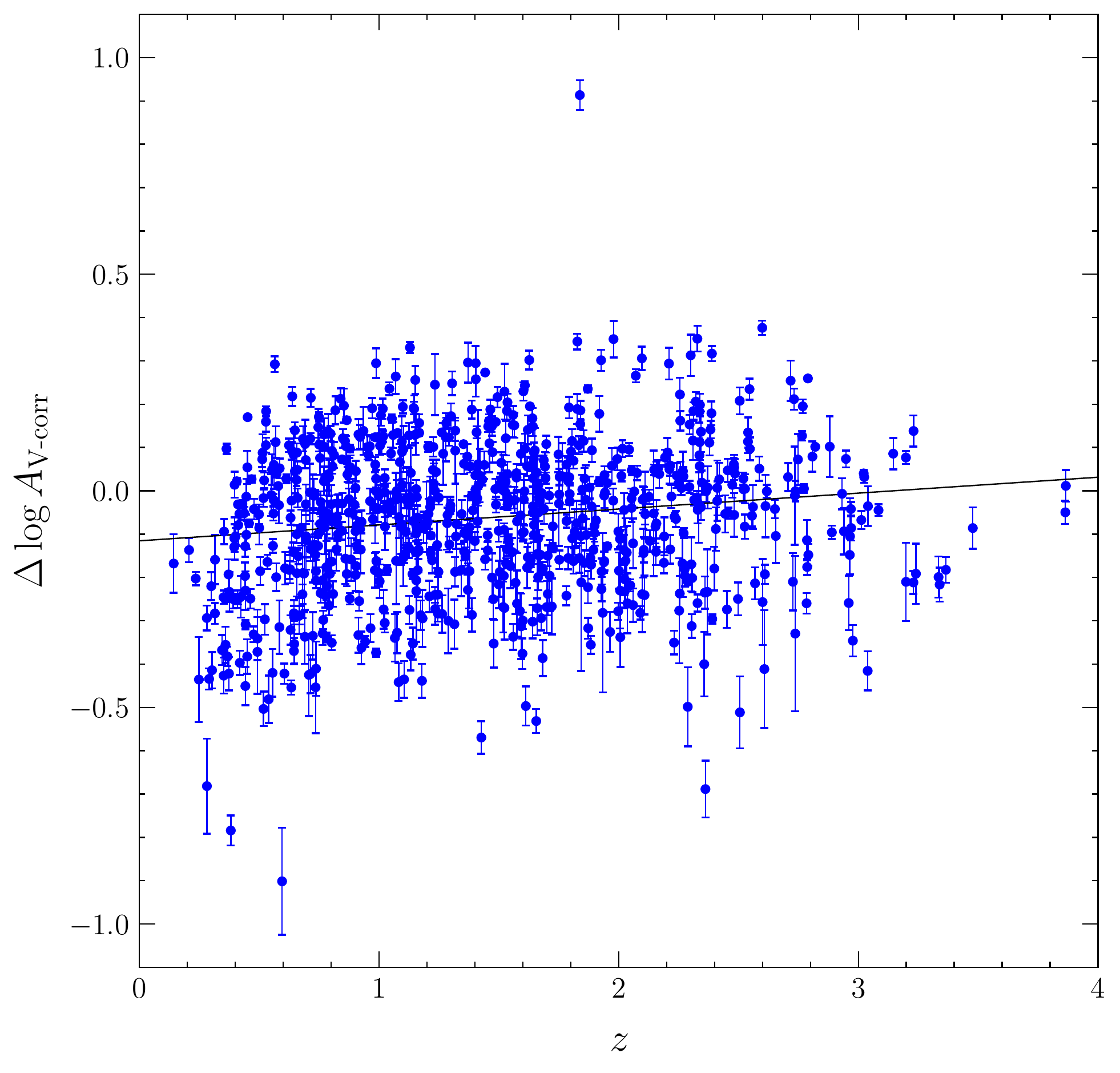}
\caption{The residuals of the $\log{A\ts{V-corr}}$$-\log{L\ts{bol}}$ relation are positively correlated with redshift, $z$. The Spearman's rank correlation coefficient is $\rho\ts{s}=0.14$, with null probability $p(>\!|\rho\ts{s}|) \simeq 10^{-4}$. The black solid line describes the best-fit relation.}

\label{fig:resid}
\end{figure}

\noindent The second term on the right-hand side of the above equation is given by the expression of Eq.\ \ref{eq:lbol}.
In this way, we can keep trace of the dependence on luminosity before searching for the evidence of a significant correlation between the residuals and $z$.
The result is shown in Fig.\ \ref{fig:resid}. We find a weak positive correlation ($\rho\ts{s}=0.14$) between the residuals and $z$, with null probability $p(>\!|\rho\ts{s}|) \simeq 10^{-4}$.
This relation can be expressed as:
\begin{equation}
    \Delta{\log{A\ts{V-corr}}}  = (-0.12\pm0.01) + (0.037\pm0.009)\,z\,.
    \label{eq:resid}
\end{equation}

\noindent A similar result has also been found in some other works in the optical \citep[e.g.][]{vandenberk2004,li2018}.
We underline that, in principle, the expression given by Eq.\ \ref{eq:resid} depends on the adopted V-correction. In Sect.\ \ref{sec:Vc} we described how such correction is connected to the spectral variability parameter, $\beta$ (see Eq.\ \ref{eq:Vcorr7}). 
In the same context, we discussed the reasons which lead us to set $\beta=1$ for all AGNs in our reference sample, as an acceptable average value. This derives from converting the wavelength-dependence of variability from \citet{morganson2014} in terms of $\beta$. The conversion returns a value which is exactly $\beta = 1.01 \pm 0.04$. If we considered the upper and lower limits of this quantity, given by the error, the above relation of Eq.\ \ref{eq:resid} would be possibly ruled out. 
On the contrary, in both cases, from our analysis it is still emerging a significant positive correlation of variability with redshift. 

\noindent The slope of the power-law SF, $\gamma$, instead, does show no significant correlation with $z$. 

In Sect.\ \ref{sec:data} we discussed that the data collected by the Catalina Surveys are unfiltered and then calibrated to Johnson's $V$ magnitude.
Since each source has been observed many times, an average optical magnitude can be estimated for all individual AGNs in the reference sample. 
Thus we compute the luminosity in the $V$ band, $L\ts{V}$ from such average magnitudes.
We find that this quantity is negatively correlated with the rest-frame variability amplitude. The Spearman's rank correlation coefficient is $\rho\ts{s}=-0.50$, with null probability $p(>\!|\rho\ts{s}|) <10^{-10}$.
They are related by the following expression:
\begin{equation*}
    \log{A\ts{V-corr}}  = (8.4\pm0.6) - (0.20\pm0.01)\,\text{log}\,L\ts{V}\,.
\end{equation*}

\noindent Furthermore, we also have estimates of the X-ray luminosity in the $0.5-4.5$ keV band for the AGNs in the reference sample. These estimates are included in the MEXSAS2 catalogue.
We find the rest-frame variability amplitude to be negatively correlated with $L\ts{X}$, as:
\begin{equation*}
    \log{A\ts{V-corr}}  = (3.5\pm0.7) - (0.10\pm0.02)\,\text{log}\,L\ts{X}\,.
\end{equation*}

\noindent Once more, this correlation is highly significant ($p(>\!|\rho\ts{s}|) < 10^{-10}$) and the Spearman's coefficient is $\rho\ts{s}=-0.26$.
These trends are in agreement with the above mentioned inverse correlation between variability amplitude and $L\ts{bol}$.
Nevertheless, the $\log{A\ts{V-corr}}$$-\log{L\ts{X}}$ relation is weaker than its optical counterpart. This is expected, since the contribution of the X-ray emission to the SED of AGNs is lower than the optical one.
In addition, we find the evidence of a clear weak correlation of the SF slope parameter $\gamma$ with both $L\ts{V}$ and $L\ts{X}$. We obtain, respectively, $\gamma = (-3.4\pm0.6) + (0.08\pm0.01)\,\log{L\ts{V}}$ and $\gamma = (-3.2\pm0.6) + (0.08\pm0.01)\,\log{L\ts{X}}$. 
The former relation is characterised by a Spearman's rank correlation coefficient $\rho\ts{s} = 0.19$, while the latter has $\rho\ts{s} = 0.20$. 
In both cases, the null probability is $p(>\!|\rho\ts{s}|) \simeq 10^{-8}$.

\begin{figure}
\centering
\includegraphics[width=\columnwidth]{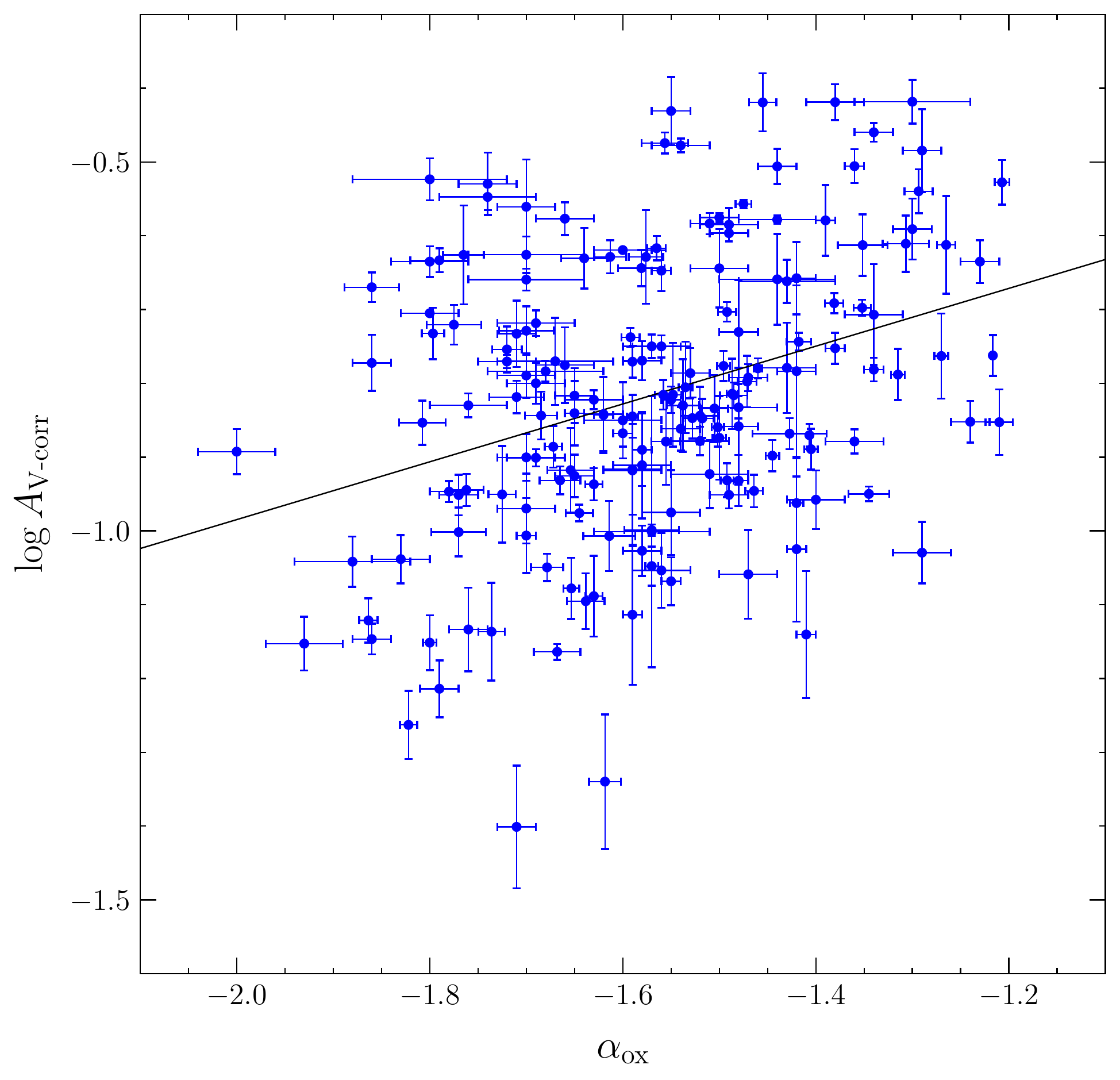}
\caption{The variability amplitude parameter, $\log{A\ts{V-corr}}$ is positively correlated with $\alpha\ts{ox}$. The Spearman's rank correlation coefficient is $\rho\ts{s}=0.29$, with null probability $p(>\!|\rho\ts{s}|) \simeq 7 \times 10^{-5}$. The black solid line describes the best-fit relation.}
\label{fig:aox}
\end{figure}

In their recent paper, \cite{chiaraluce2018} did investigate the X-ray/UV ratio of a subsample of 636 AGNs originally included in the MEXSAS2 catalogue.
Their sample is then complemented with measurements of the $\alpha$\ts{ox} parameter for each individual source.
From a simple match, it is emerging that solely 178 of such AGNs are also present in our reference sample.
Thus, we aim to analyse the dependence of the SF variability parameters on $\alpha\ts{ox}$ for these objects.
We find a significant positive correlation (see Fig.\ \ref{fig:aox}) between variability amplitude and $\alpha\ts{ox}$, which can be expressed as follows:
\begin{equation*}
    \log{A\ts{V-corr}}  = (-0.2\pm0.1) + (0.39\pm0.08)\,\alpha\ts{ox}\,.
\end{equation*}

\noindent The Spearman's rank correlation coefficient is $\rho\ts{s}=0.29$, with null probability $p(>\!|\rho\ts{s}|) \simeq 7 \times 10^{-5}$.
We underline that this result, together with the previously discussed $\log{A\ts{V-corr}}-$$\log{L\ts{V}}$ and $\log{A\ts{V-corr}}-$$\log{L\ts{bol}}$ negative correlations, is expected also in the light of the frequently reported inverse correlation between $\alpha\ts{ox}$ and optical/UV luminosity \citep[e.g.][]{vignali2003,gibson2008, vagnetti2010, chiaraluce2018}.
On the contrary, we do not find the evidence of a significant correlation between the SF slope parameter $\gamma$ and $\alpha\ts{ox}$.

\section{Comparison with X-ray variability}
\label{sec:compX}
Our AGN sample has been chosen to investigate the optical variability properties of the sources included in the MEXSAS2 catalogue. On the other hand, MEXSAS2 was introduced in the context of an X-ray variability analysis.
Thus we think it is interesting to compare variability in these two bands, in order to obtain a wider and deeper knowledge of the sample.
A fair comparison needs us to recall that in the X-ray, due to the lack of a large number of observations of individual sources, we have to consider the ensemble SF.
Then, it is necessary to search for possible relations between optical and X-ray ensemble SFs.  
We choose to classify the objects in the reference sample in terms of their variability parameters, that is, amplitude and slope of the power-law SF.
\begin{figure*}
\vspace{-0.5cm}
\centering
\includegraphics[width=\columnwidth]{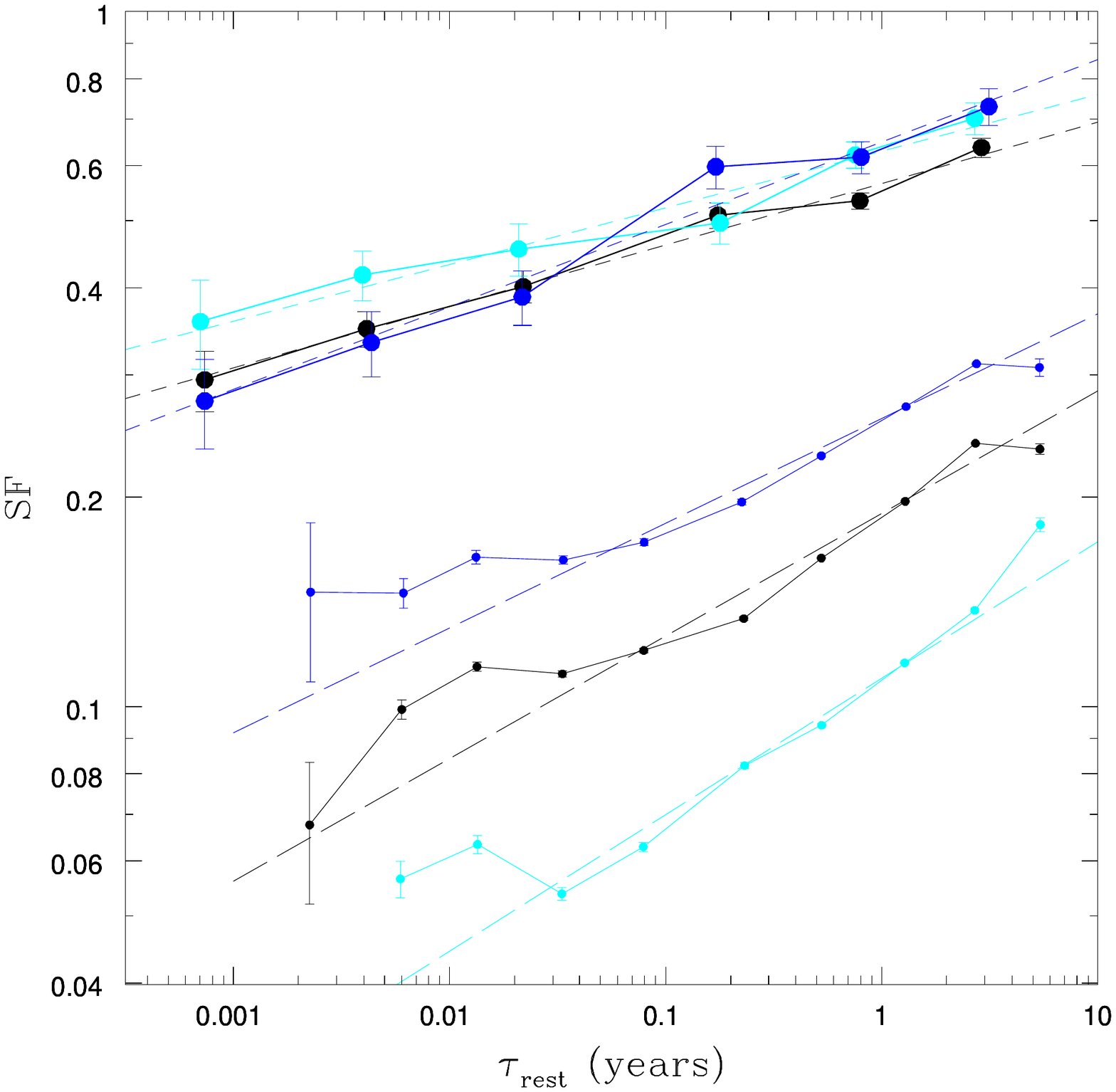}
\includegraphics[width=\columnwidth]{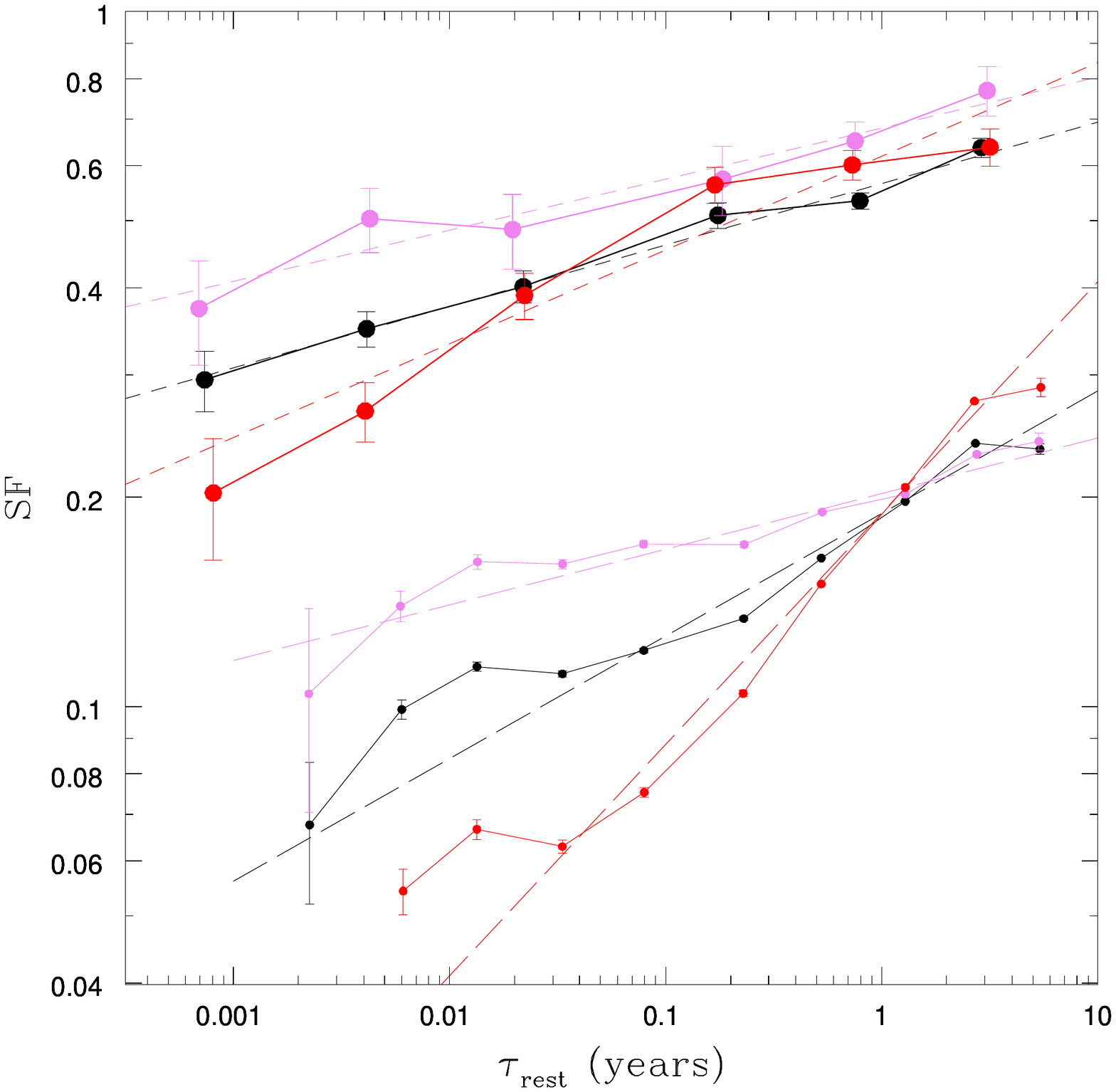}
\vspace{-2cm}
\caption{\emph{Left panel}: comparison between optical and X-ray ensemble SFs associated to the AGN subsamples with larger (in blue) and smaller (in cyan) optical variability amplitude. \emph{Right panel}: comparison between optical and X-ray ensemble SFs associated to the AGN subsamples with steeper (in red) and shallower (in pink) optical individual power-law SF. In both panels thick and thin solid lines and dots describe, respectively, the X-ray and optical ensemble SFs. Short and long dashed lines refer, similarly, to the least-squares fit to the X-ray and optical ensemble SFs. The fits are weighted for the number of lag values included in each available bin. We underline that each SF is corrected for the contribution of the photometric errors. V-correction is also implemented as well. X-ray ensemble SFs are computed as 2.5 times Eq.\ \ref{eq:sfx}, to convert the logarithms of fluxes into magnitudes. 
As a benchmark, we show (in black) the optical and X-ray ensemble SFs related to the whole reference sample.}
\label{fig:sfcompX}
\end{figure*}
In practice, we make a distinction between the AGNs whose SF has larger ($\log{A}\ts{V-corr} - \sigma_{\log{A_V}} > -0.812$) and smaller ($\log{A}\ts{V-corr} + \sigma_{\log{A_V}} < -0.812$) values of variability amplitude.
A similar approach is carried out to distinguish between those AGNs with steeper ($\gamma - \sigma_\gamma > 0.167$) and shallower ($\gamma + \sigma_\gamma < 0.167$) power-law SF. 
The above thresholds take into account the uncertainties on the variability parameters, $\sigma_{\log{A_V}}$ and $\sigma_\gamma$ to remove objects that could be included in both subsamples, within their errors. 
These thresholds are arbitrary and they tend to distribute the whole sample in two almost equally populated subgroups.
Considering two sufficiently well populated subsamples is useful to compute consistent ensemble SFs, especially in the X-ray where the observations of each individual source are fewer than in the optical.
In addition, we recall that the X-ray SFs are computed from flux measurements, as follows:
\begin{equation}
\label{eq:sfx}
\text{SF}(\tau) \equiv \sqrt{\langle[\log{f\ts{X}}(t+\tau) - \log{f\ts{X}}(t)]^2\rangle - \sigma^2_\text{noise}}\,,
\end{equation}

\noindent where the contribution of the photometric uncertainties, $\sigma^2_\text{noise}$ has the same meaning of that of Eq.\ \ref{eq:sf}, once the errors on magnitudes at each epoch are replaced by those on the logarithms of fluxes at the corresponding epochs.

Hence, a fair comparison between the X-ray and optical ensemble SFs requires the former to be shifted upwards by a factor of 2.5.
In Fig.\ \ref{fig:sfcompX}, where we show the results of our analysis, the X-ray ensemble SFs are always above their optical counterparts. This trend does agree with previous determinations of the ensemble SF variability amplitude in the optical \citep[e.g.][]{morganson2014,li2018} as well as in the X-ray \citep[e.g.][]{vagnetti2016}.
First, let us focus on the left panel of Fig.\ \ref{fig:sfcompX}. In this case we distinguish between those subsamples of AGNs with larger and smaller optical variability amplitude, whose ensemble SFs are coloured respectively in blue and cyan.
We can notice that the X-ray ensemble SF shows almost no sensitivity with respect to these AGN samples, that is, X-ray variability amplitude appears to be uncorrelated with the optical.
An approximately similar result has been found by \citet{edelson2019} in a small sample of four intensively observed AGNs. Though they find a substantial lack of correlation between optical and X-ray bands, they suggest the evidence of a strong correlation in the optical/UV variability.
To complicate the scenario even further, a recent paper by \citet{xin2020} does confirm this strong correlation in a much deeper sample of $\sim1300$ AGNs but also rules out at $\sim95\%$ confidence level that the intrinsic UV and optical variations of all quasars are fully correlated.
They then suggest the existence of physical mechanisms that can generate uncorrelated optical and UV flux variations, such as expected, for example, from local temperature fluctuations \citep[e.g.][]{kawaguchi1998, trevese2002, dexter2011}.

However, coming back to our result, we also need to take into account that the optical emission from the disk is supposed to be reprocessed by the high-energy photons produced by the inverse-Compton (IC) scattering in the corona.
In this lamp-post model \citep[e.g.][]{epitropakis2016} one would probably expect an intimate connection between X-ray and optical flux variations.
On the other hand, the interconnection between these optical/X-ray variations depends also on the black hole mass \citep[e.g.][]{uttley2006}. In our reference sample the majority (95\%) of AGNs hosts quite massive central black holes with $10^8\lesssim M\ts{BH}/M_\odot \lesssim 10^{10}$.
For these sources, the region of the disk associated to the emission in the $V$ band lies at a distance between $\sim 1$ light-week and few light-months from the centre. 
At such distances the X-ray radiation of the corona may possibly only partially affect the intrinsic emission of that portion of the disk. This would probably cause X-ray flux variations, which are normally supposed to drive the optical ones \citep[e.g.][]{krolik1991}, to play a more marginal role, once reprocessed in the disk, in the variability of the $V$ band.
This effect may be also explained in terms of the disk reprocessing model by \citet{gardner2017}. This model predicts that the hard X-ray photons do not directly illuminate the accretion disk. Indeed, they are thought to interact with an intermediate structure, namely a Comptonised disk, which obscures the inner X-ray emitting region. The source of flux illuminating the outer disk, that is the main optical/UV emitter, is then such Comptonised disk. Thus, it is possible that the amplitude of the X-ray variations is smeared out through dissipation processes occurring inside this structure, before reaching the outer regions of the disk, where their role in contributing to the optical/UV variations may be somehow limited. 
In this way, one can justify the apparent lack of correlation between the optical and X-ray variability amplitudes.

On the contrary, the right panel of Fig. \ref{fig:sfcompX} seems to suggest a connection between these two bands. Indeed, in this figure we can observe the ensemble SFs related to the subsamples with steeper (in red) and shallower (in pink) optical power-law SF.
This can be a hint that those AGNs with steeper SFs in the optical do also present steeper X-ray SFs, and vice versa.
Such result needs to be confirmed by deeper investigations and, at present, we can only formulate hypotheses.
A viable option would be that invoking the propagating fluctuations model, which assumes local fluctuations of the accretion rate in the disk \citep[][]{lyubarskii1997}. These fluctuations are supposed to propagate inward towards the central regions. During their flow they are modulated according to the local viscous timescale, which, depending on the radius, becomes progressively shorter. 
The inner regions, in this way, are sensitive to what happens outwards.
If the propagation process is such that the given fluctuation transfers its imprint on the following region, one would possibly justify the result in the right panel of Fig.\ \ref{fig:sfcompX}.
In this scenario, if the region of the disk responsible for the emission in the $V$ band is accumulating variability over increasing time lags at a certain rate, the innermost regions, including the corona, should keep trace of such behaviour.
As a consequence, AGNs with steeper (or shallower) SFs in the optical do present the same trend in the X-ray.

\section{Summary and conclusions}
\label{sec:summ}
The Catalina Surveys have carried out a large number of multi-epoch observations of many AGNs. We used these data to investigate the individual optical variability properties of a sample of 795 AGNs, extracted from the MEXSAS2 catalogue \citep{serafinelli2017}.
We chose the structure function as variability estimator, which works in the time domain and represents a popular tool to analyse the flux variations of AGNs. Our sample was also complemented with measurements of physical quantities which we aimed to relate to the variability parameters obtained from the SF, once modelled in terms of a power-law as $\text{SF}(\tau) = A\,(\tau/\tau_0)^\gamma$. 
Our main results may be summarised as follows. 
\begin{enumerate}
    \item{In order to account for the dependence of variability on the emission frequency, we have introduced the V-correction \citep[][]{vagnetti2016} as a simple tool to correctly quantify the amount of variability in the rest frame of each source (see discussion in Sect.\ \ref{sec:Vc}).}
    \item{We have found the evidence of a strong inverse correlation between variability amplitude and bolometric luminosity. Such trend is compatible with previous works \citep[e.g.][]{vandenberk2004, macleod2010, morganson2014}.
    In particular, our result suggests that $L\ts{bol}$ may be the main driver of AGN variability, since the dependencies of variability amplitude on both black hole mass and Eddington ratio appear to be much weaker, in analogy with, e.g., \citet{caplar2017}.
    Rest-frame variability amplitude, $\log{A\ts{V-corr}}$, appears to be inversely correlated also with the optical luminosity in the $V$ band as well as with the X-ray luminosity, where the dependence on the latter is weaker. 
    We have seen that the residuals of the $\log{A\ts{V-corr}}-\log{L\ts{bol}}$ relation are significantly correlated with redshift, a result which is in agreement with some previous studies in the optical \citep[e.g.][]{li2018}.
    Though this trend is not that strong, it may suggest that AGNs were possibly more active in the early universe, thus eventually indicating an evolutionary property of their variability.  
    In addition, we have also obtained the evidence of a significant correlation between $\log{A\ts{V-corr}}$ and $\alpha\ts{ox}$ (see Sect.\ \ref{sec:corr}).}
    \item{The rich sampling of the Catalina Surveys has allowed us to compute individual SFs. Hence, it has been possible to relate also their resulting slope, $\gamma$, with physical quantities. We have obtained the indication of a weak positive correlation of $\gamma$ with the bolometric luminosity $L\ts{bol}$ and also, albeit marginally, with the black hole mass $M\ts{BH}$.}
    \item{We have used the data included in MEXSAS2 to compare the optical and X-ray ensemble variability properties of the AGNs in our reference sample. First, we have divided these objects between those with larger and smaller values of optical variability amplitude at a rest-frame time lag of 1 yr. We have found that both of these subsamples have little impact on the X-ray ensemble SFs related to the same sources. Our result suggests that X-ray variability amplitude may be uncorrelated with the optical. 
    In a similar way, we have divided the reference sample into those AGNs with steeper and shallower power-law SFs. In this case, we have found a connection between these subsamples.
    Indeed, we have obtained an indication that those AGNs with steeper SFs in the optical do present the same trend in the X-ray, and vice versa.} 
\end{enumerate}

\noindent With this work we aim to shed light on the importance of richly-sampled photometric campaigns, which are fundamental to investigate individual variability features of AGNs.
These campaigns should be carried out, regardless of the given window of the electromagnetic spectrum we are looking at.
Indeed, combining multi-wavelength observations may result in a better understanding of the physical mechanisms related to AGNs, and variability studies represent useful probes for such investigation.

As shown in the previous sections, optical/UV and X-ray bands are intimately connected. These bands do carry information about the interplay of two physical regions, the accretion disk and the corona.
Since in both cases their variability spans wide ranges of characteristic timescales, not only need the above campaigns to be performed with repeated sampling, but they should also be sufficiently extended in time to cover from short to long-term variability. 
In addition, it is clear that very accurate photometry is mandatory. 
Photometric uncertainties such as those of the Catalina Surveys may limit to some extent the reachable goals of a variability study. Nevertheless, once some arrangements are done, we have shown that a meaningful analysis can still be carried out.

In the near future, new facilities such as Large Synoptic Survey Telescope \citep[LSST;][]{ivezic2019} will be operating and we expect very accurate photometric data to be obtained. This will open an unprecedented opportunity to unveil AGN variability properties in detail.

\vspace{1cm}

\section*{Acknowledgements}
We thank the anonymous referee for the useful comments that helped us improving the clarity of the present work. 
This work has made use of data obtained from the Catalina Surveys.
The CSS survey is funded by the National Aeronautics and Space Administration under Grant No.\ NNG05GF22G issued through the Science Mission Directorate Near-Earth Objects Observations Program. The CRTS survey is supported by the U.S.~National Science Foundation under grants AST-0909182 and AST-1313422.
M.\ L.\ acknowledges the financial support of MIUR (Ministero dell'Istruzione, dell'Università e della Ricerca) to the Ph.D. programme in Astronomy, Astrophysics and Space Science.
F.\ V.\ and M.\ P.\ acknowledge the financial contribution from the agreement ASI-INAF n.2017-14-H.O.
R.\ M.\ acknowledges the financial support of INAF (Istituto Nazionale di Astrofisica), Osservatorio Astronomico di Roma, ASI (Agenzia Spaziale Italiana) under contract to INAF: ASI 2014-049-R.0 dedicated to SSDC.

\section*{Data availability}

The data underlying this article are available in the article and in its online supplementary material.




\bibliographystyle{mnras}
\bibliography{biblio} 




%
%


\bsp	
\label{lastpage}
\end{document}